\documentclass{article}
\usepackage{latexsym}
\usepackage{amsmath, amssymb, amsthm, bm}
\usepackage{hyperref}
\usepackage{graphicx}
\usepackage{multirow}
\usepackage{natbib}
\usepackage{boxedminipage}

\usepackage{tikz}
\usetikzlibrary{cd}
\usepackage{footnote}

\DeclareMathOperator*{\argmin}{arg\,min}
\DeclareMathOperator*{\argmax}{arg\,max}

\hypersetup{
	colorlinks=true,
    citecolor=black
}

\begin{document}
\setcounter{page}{0}
\title{\bf State Space Emulation and Annealed Sequential Monte Carlo for High Dimensional Optimization}
\author{Chencheng Cai and Rong Chen\footnote{
Rong Chen is Professor at Department of
    Statistics, Rutgers University, Piscataway, NJ 08854. E-mail:
    rongchen@stat.rutgers.edu. Chencheng Cai is a Ph.D student
    at Department of Statistics, Rutgers
    University, Piscataway, NJ 08854. E-mail:
    chencheng.cai@rutgers.edu.
Rong Chen is the
    corresponding author. Chen's research is supported
in part by National Science Foundation
grants DMS-1503409, DMS-1737857 and IIS-1741390.
}
\\ \vspace{0.2cm} {Rutgers University}\\}
\date{}
\maketitle

%\date{}

\begin{abstract}
Many high dimensional optimization problems can be reformulated into a problem of finding the optimal state path under an equivalent state space model setting. In this article, we present a general emulation strategy for developing a state space model whose likelihood function (or posterior distribution) shares the same general landscape as the objective function of the original optimization problem. Then the solution of the optimization problem is the same as the optimal state path that maximizes the likelihood function or the posterior distribution under the emulated system. To find such an optimal path, we adapt a simulated annealing approach by inserting a temperature control into the emulated dynamic system and propose a novel annealed Sequential Monte Carlo (SMC) method that effectively generating Monte Carlo sample paths utilizing samples obtained on the high temperature scale. Compared to the vanilla simulated annealing implementation, annealed SMC is an iterative algorithm for state space model optimization that directly generates state paths from the equilibrium distributions with a decreasing sequence of temperatures through sequential importance sampling which does not require burn-in or mixing iterations to ensure quasi-equilibrium condition. Several applications of state space model emulation and the corresponding annealed SMC results are demonstrated.
\end{abstract}

\newpage
\section{Introduction}
High dimensional global optimization algorithms are being widely investigated since more and more applications involve high dimensional complex data nowadays. The gradient descent algorithm and its variations \citep{bertsekas1997nonlinear} require the objective function to be convex or uni-modal so that the found local optimal is global.
Recent research in machine learning involves many non-convex optimization problems \citep{anandkumar2014tensor,arora2012computing, netrapalli2014non, agarwal2014learning}. However, many non-convex problems remain NP-hard and the theory is only available for their convex relaxations \citep{jain2017non}.
Deterministic optimization algorithms \citep[for example, ][]{Hooke1961direct, nelder1965simplex, Land1960An} may result in certain type of exhaustive search, which is computationally expensive in a high dimensional space. Stochastic optimization algorithms utilizes Monte Carlo simulations to explore the parameter space in a stochastic and often more efficient way \citep{kiefer1952stochastic, Kirkpatrick1983Optimization, mei2018mean}.
\par
In this article, we propose an emulation approach that reformulates a high dimensional optimization problem into the problem of finding the most likely state path problem in a state space model.
The state space models is a class of models that describes the behavior of a usually high-dimensional random variable in a form of dynamic evolution,
with wide applications in mathematics, physics and many other fields. Many high-dimensional optimization problems can be transformed to finding the optimal state path under an equivalent state space model, whose likelihood function shares the same general landscape as the objective function of the original optimization problem. 
To be more specific, for a high-dimensional optimization problem with the objective function $f(x)$, we construct an emulated state space model whose likelihood function is proportional to a Boltzmann-like distribution $\exp(-\kappa f(x))$, where $\kappa>0$ is the inverted temperature. 

There are several existing heuristic approaches using the emulation idea.
\cite{cai2009variable} transforms a regression variable selection problem with many predictors into an optimization problem over the high dimensional binary space $\{0, 1\}^p$, which can be further converted to the most likely path problem in a state space model with binary-valued states indicating the variable selection, even though the predictors have no chronological order in nature. 
% \cite{cai2009variable} shows that with the idea of look-ahead strategy borrowed from SMC, the performance of their approach can be significantly boosted. 
\cite{kolm2015multiperiod} reformulates a portfolio optimization problem to a state space model by mapping the utility function to the log-likelihood function. The utility function is then optimized through finding the most likely path in the corresponding state space model by applying the Viterbi algorithm \citep{viterbi1967error} over Monte Carlo samples. Similarly, \cite{irie2016bayesian} relates the multi-period portfolio optimization problem to the log-likelihood of a mixture of linear Gaussian dynamic systems and proposed an algorithm based on the Kalman filter \citep{kalman1960new} and EM algorithm \citep{dempster1977maximum} to find the most likely path. 

These studies map high dimensional optimizations to a problem under state space model settings. However, it remains a challenging problem to find the most likely path analytically and numerically.
%without converting it back to the optimization problem that is similar to the original one. 
For example, the approach in \cite{cai2009variable} is difficult to be generalized to a continuous space. The Viterbi algorithm used in \cite{kolm2015multiperiod} requires the dynamic system to be Markovian and non-singular and it needs a large sample size in general to achieve high accuracy. The combination of Kalman filter and EM algorithm proposed in \cite{irie2016bayesian} works only when the underlying distribution can be represented as a mixture of Gaussian distributions. 

In this paper, we propose a new Sequential Monte Carlo (SMC) based simulated annealing approach, named annealed SMC, to find the most likely path in a state space model. 
The SMC algorithm is a class of Monte Carlo methods that draws samples from the state space model systems in a sequential fashion. With the sequential importance sampling and resampling (SISR) scheme, SMC is extremely powerful in sampling from complex dynamic systems, especially for the state space models \citep{gordon1993novel, kitagawa1996monte,kong1994sequential, liu1995blind, liu1998sequential, pitt1999filtering, chen2000adaptive, doucet2001introduction}. 
For a high-dimensional optimization problem with the objective function $f(x)$, we construct an emulated state space model whose likelihood function is proportional to a Boltzmann-like distribution $\exp(-\kappa f(x))$, where $\kappa$ is the inverted temperature.
To mimic the (physical) annealing procedure in a non-interactive, non-quantum thermodynamic system \citep{Kirkpatrick1983Optimization}, we choose a sequence of decreasing temperatures $\kappa_0 < \kappa_1 <\dots<\kappa_K$, which corresponds to a sequence of emulated state space models. 

We start from drawing sample paths from the base emulated state space model at a high base temperature $\kappa_0$. Samples from a low temperature (large $\kappa$) system are close to the optimal sample path since the distribution is sharp at a low temperature, but 
drawing from such a distribution directly is usually difficult. With annealed SMC, samples of a low temperature system can be obtained by utilizing samples obtained at higher temperature. Eventually, all the SMC sample paths converge to the most likely one. The sequence of temperatures $\kappa_0 < \kappa_1 <\dots<\kappa_K$ provides a slow-changing path from the base emulated state space model at $\kappa_0$, which is easy to sample from but not very useful for optimization, to the target emulated state space model at $\kappa_K$, which is difficult to sample but provides solutions to the optimization problem. 

Four examples of state space emulation and their corresponding simulation experiments will be demonstrated. The smoothing spline optimization problem \citep{green1993nonparametric} is emulated by a state space model in which the function value and its first two derivatives of the spline function at every knot are formulated as a 3-dimensional vector AR process. A regularized regression problem with $p$ covariates, such as the ridge regression and LASSO \citep{tibshirani1996regression} setting, is transformed into a state space model of length $p$, where the $p$ coefficients form the hidden states. The $\ell$1 trend filtering problem \citep{kim2009ell_1} is emulated by an AR(2) model with Laplace evolutions. In constructing an optimal stock trading strategy with transaction cost consideration \citep{kolm2015multiperiod}, the hidden states of the emulated system consist of stock positions during the period of consideration, and the profit/loss is formulated into the observation part and the transaction costs are taken into state dynamics. 

The rest of the paper is organized as follows. Section \ref{sec:emulation} first briefly reviews state space models then introduces the principles of state space emulation. Four emulation examples are provided in Section \ref{sec:emulation-example}. Section \ref{sec:asmc} introduces the framework of annealed SMC designed to find the most likely path. The practical details of using the algorithm are discussed. Simulation results corresponding to the four examples in Section \ref{sec:emulation-example} are shown in Section \ref{sec:simulation}. Section \ref{sec:conclusion} concludes.

\section{State Space Model and State Space Emulation}\label{sec:emulation}
\subsection{State Space Model}
State space model is a class of models for describing the mechanism of a sequence of observations $\bm y_T = (y_1, \dots, y_T)$ with a sequence of latent variables $\bm x_T=(x_1,\dots, x_T)$. The latent variables $\bm x_T$ are assumed to follow a discrete-time stochastic process governed by the state equations
\begin{equation}
    p(x_t\mid \bm x_{t-1}) = p_t(x_t\mid \bm x_{t-1}), \label{eq:state-equation}
\end{equation}
for $t = 2,\dots, T$, and $x_1$ follows its marginal distribution $p_1(x_1)$.
When the distribution of $x_t$ conditioned on $\bm x_{t-1}$ does not depend on $\bm x_{t-2}$ such that $p(x_t\mid \bm x_{t-1}) = p(x_t\mid x_{t-1})$, the system is Markovian. 
The observations $\bm y_{T}$ are generated conditionally independently according to the corresponding latent variables through the observational equations
\begin{equation}
    p(y_t\mid x_t) = g_t(y_t\mid x_t),\label{eq:observation-equation}
\end{equation}
for $t = 1,\dots, T$.
In inference problems, the formulas of the state equations $p_t(\cdot)$ and the observation equations $g_t(\cdot)$ are usually known except a set of  unknown parameter of interest $\theta$. In this paper, we assume $p_t(\cdot)$ and $g_t(\cdot)$ are completely known with no unknown parameters, and we are interested in inference on the latent states $\bm x_T$. Estimating $\bm x_T$ from the observations $\bm y_T$ under the likelihood 
principle is known as the most likely path problem in hidden Markov models.

In terms of estimating $\bm x_T$, the state equations provide the prior information 
\begin{equation}
    \pi(\bm x_T)\propto p_1(x_1)\prod_{t = 2}^T p_t(x_t\mid \bm x_{t-1}), \label{eq:prior}
\end{equation}
and the observation equations serve as the likelihood functions
\begin{equation}
    p(\bm y_T\mid \bm x_T) = \prod_{t=1}^T g_t(y_t\mid x_t).\label{eq:likelihood}
\end{equation}
A maximum-a-posterior (MAP) estimator can be obtained by maximizing the posterior function in (\ref{eq:posterior}).
\begin{equation}
    \pi(\bm x_T|\bm y_T)\propto p_1(x_1)g_1(y_1\mid x_1)\prod_{t = 2}^T p_t(x_t\mid \bm x_{t-1})g_t(y_t\mid x_t).\label{eq:posterior} 
\end{equation}

When both $p_t(\cdot)$ and $g_t(\cdot)$ are Gaussian, the maximum of (\ref{eq:posterior}) can be obtained easily using Kalman filter and smoother \citep{kalman1960new}. In general cases when an analytic solution to optimize (\ref{eq:posterior}) is infeasible, the MAP estimator can be obtained by drawing sample paths $\{(x_1^{(i)}, \dots, x_T^{(i)})\}_{i=1,\dots, n}$ from the posterior distribution $(\ref{eq:posterior})$. We will discuss details in estimating most likely path using Monte Carlo methods in Sections \ref{sec:mlp} and \ref{sec: asmc-asmc}.

\subsection{State Space Emulation}\label{sec:emulation-sse}
We propose a state space emulation approach for solving high dimensional optimization problems.
The approach constructs a state space model so that the original optimization problem is equivalent to finding the most likely state path under the state space model. 

Let $f:\mathcal X^d\rightarrow\mathbb R$ be the objective function to be minimized and $\xi:\mathbb R\rightarrow\mathbb [0,+\infty)$ be a monotone decreasing function. Then minimizing $f(x)$ is equivalent to maximizing $\phi(x):=\xi(f(x))$ such that
$$\argmin_{x\in\mathcal X^d} f(x) = \argmax_{x\in\mathcal X^d} \phi(x).$$
% To be more specific, let $(\mathcal X, \mathcal F_x)$ be a measurable space, where $\mathcal F_x$ is all Borel sets over $\mathcal X$ when $\mathcal X$ is a continuous random variable and $\mathcal F_x$ is the power set of $\mathcal X$ when $\mathcal X$ is discrete. Suppose $f: \mathcal X^d \rightarrow \mathbb R$ is a $\mathcal F_x^d$-measurable objective function to be minimized. 
% If a measurable function $\xi:\mathbb R\rightarrow [0, +\infty)$ is monotonely decreasing and bounded, then 
% $$\argmin_{x\in\mathcal X^d} f(x) = \argmax_{x\in\mathcal X^d} \xi(f(x)),$$
% provided the left hand side exists and is unique. 
% Furthermore, let $\mu$ be the Lebesgue measure if $\mathcal X$ is continuous or the counting measure if $\mathcal X$ is discrete.
% If $\phi = \xi\circ f$ is integrable with respect to measure $\mu$ such that $\int \phi(x) d\mu(x) < \infty$, then $(\mathcal X^d, \mathcal F_x^d, P_\phi)$ is a valid probability space, where
% $$
% P_\phi (A) = \dfrac{\int_A \phi(x)d\mu(x)}{\int_{\mathcal X^d} \phi(x)d\mu(x)},
% $$
% for any measurable set $A\in\mathcal F_x^d$, and $\phi(x)$ is the corresponding (un-normalized) probability density/mass function. 
Furthermore, if there exists a state space model whose posterior function (\ref{eq:posterior}) is proportional to $\phi(x)$ such that
$$\pi(\bm x_T\mid \bm y_T)\propto \phi(\bm x_T)=\xi(f(\bm x_T)),$$
with artificially designed state equations $\{p_t(\cdot)\}_{t= 1:d}$, and observation equations $\{g_t(\cdot)\}_{1:d}$, we call the state space model an ``emulated" state space model. The observations $\bm y_T$ can be either certain observations involving in the original optimization problem (e.g. the observed points in the smoothing spline problem in Section \ref{sec:emulation-example-spline}) or artificially designed. Note that it is always possible to rewrite any joint distribution function $\phi(\bm x_T)$ in the form of (\ref{eq:prior}) as
$$\phi(\bm x_T) = \phi(x_1,\dots, x_T) = \phi_1(x_1)\prod_{t=2}^T\phi_t(x_t\mid \bm x_{t-1}),$$
where $\phi_1(x_1)=\int_{\mathcal X^{t-1}} \phi(\bm x_T)dx_2\dots dx_T$ and $\phi_t(x_t\mid \bm x_{t-1})=\dfrac{\int_{\mathcal X^{T-t}}\phi(\bm x_T)dx_{t+1}\dots d_T}{\int_{\mathcal X^{T-t+1}}\phi(\bm x_T)dx_{t}\dots d_T}$. Often such a series of conditional distribution is difficult to sample from or to be evaluated. However, in certain problems as our examples shown later, it is possible to reformulate the conditional distribution to $\phi_t(x_t\mid \bm x_{t-1})= p_t(x_t\mid \bm x_{t-1})g_t(y_t\mid x_t)$, in which $p_t(x_t\mid \bm x_{t-1})$ is easy to generate sample from and $g_t(y_t\mid x_t)$ is easy to be evaluated, for some designed $y_t$.  
Minimizing the objective function is then the same as finding the most likely path for the emulated state space model. The emulated state and observation equations provide guidance for annealed SMC implementation, even though they are artificial.

A common choice for the function $\xi(\cdot)$ is the Boltzmann distribution function 
\begin{equation}
\xi(s) = e^{-\kappa s},\label{eq:boltzmann}
\end{equation}
where $\kappa$ is a positive constant that relates to the temperature in statistical physics. 
In statistics, the Boltzmann function in (\ref{eq:boltzmann}) links the least square method to the maximum likelihood approach with i.i.d. Gaussian noise.
With this choice of $\xi(\cdot)$, the system has a physical interpretation: The objective function $f(\cdot)$ is regarded as the possible energy levels in a non-quantum thermodynamic system. Assuming no interactions, the number of particles at the energy $f(x)$ follows the Boltzmann distribution under thermodynamic equilibrium. The integrability of $\phi(x)$ ensures the existence of the canonical partition function such that this physical canonical system is valid. The minimization of $f(\cdot)$ is now equivalent to find the base energy level, which inspires the use of simulate annealing of this thermodynamic system. More details will be discussed in Section \ref{sec:asmc}.

\subsection{Examples}\label{sec:emulation-example}
\subsubsection{Cubic Smoothing Spline}\label{sec:emulation-example-spline}
Consider a nonparametric regression model 
$$y_t = m(x_t) + \epsilon_t$$
with equally spaced $x_t$. Without loss of generality, let $x_t=t$ and treat it as time.\\
\par
The cubic smoothing spline method \citep{green1993nonparametric} estimates a continuous function $m(t)$ by minimizing
\begin{equation}
    \sum_{t=1}^T (y_t - m(t))^2 + \lambda\int \left[m''(t)\right]^2dt.\label{eq:smoothingspline-objective}
\end{equation}
% given a sequence of observations $y_1, \dots, y_T$ at equally spaced locations. 
The first term in (\ref{eq:smoothingspline-objective}) is the total squared tracking errors at the observation times and the second term is the penalty term on the smoothness of the latent function $m(\cdot)$, where $\lambda$ controls the regularization strength. 
% Without loss of generality, we mark the locations as $1,2,\dots, T$, and we can treat them as time.
Given the values of $m(1),\dots, m(T)$, the minimizer of the second term is a natural cubic spline that interpolates $m(1),\dots, m(T)$ (see \cite{green1993nonparametric}). Hence, the solution to minimize (\ref{eq:smoothingspline-objective}) is a natural cubic spline, which is second-order continuously differentiable and is a cubic polynomial in all intervals $[t, t+1]$ for $t = 1,\dots, T-1$ and is linear outside $[1,T]$.

Define the derivatives of $m(t)$ at each observation at time t as
\begin{align*}
    a_t  = m(t),\quad
    b_t  = m'(t),\quad
    c_t  = m''(t)/2,\quad
    d_t  =\lim_{s\rightarrow t_-} m'''(s) / 6.
\end{align*}

By the constraints of natural cubic spline, we have the following recursive relationships:
\begin{align*}
    a_{t+1} &= a_t + b_t + c_t + d_{t+1},\\
    b_{t+1} &= b_t + 2c_t + 3d_{t+1},\\
    c_{t+1} &= c_t + 3d_{t+1},
\end{align*}
with $c_1=c_T=0$.
Furthermore, by substituting $d_{t+1}$ with  $(c_{t+1} - c_t)/3$ in the expressions of $a_t$ and $b_t$, we have
\begin{align}
    a_{t+1} &= a_t + b_t + (c_{t+1} + 2c_t)/3,\label{eq:ss-recursive-a}\\
    b_{t+1} &= b_t + c_t + c_{t+1}.\label{eq:ss-recursive-b}
\end{align}
We will use the recursive relationships in (\ref{eq:ss-recursive-a}) and (\ref{eq:ss-recursive-b}) for the construction of state space emulation.
With this notation, the second term in (\ref{eq:smoothingspline-objective}) can be expended as 
$$\lambda\int \left[m''(t)\right]^2dt = \lambda\sum_{t=1}^{T-1}\int_t^{t+1} \left[6(s-t)d_{t+1}+2c_t\right]^2ds = \dfrac{4}{3}\lambda\sum_{t=1}^{T-1}(c_t^2+c_tc_{t+1}+c_{t+1}^2).$$
In this case, the original optimization problem (\ref{eq:smoothingspline-objective}) over all second order differentiable functions becomes minimizing
\begin{equation}
    f(\bm x_T) = \sum_{t=1}^T (y_t - a_t)^2 + \dfrac{4}{3}\lambda\sum_{t=1}^{T-1}(c_t^2+c_tc_{t+1}+c_{t+1}^2),\label{eq:smoothingspline-objective-ssm}
\end{equation}
where $\bm x_T = \{(a_t, b_t, c_t)\}_{t=1,\dots, T}$ satisfies the recursive relationships (\ref{eq:ss-recursive-a}) and (\ref{eq:ss-recursive-b}) and the boundary condition $c_1=c_T=0$. Note that $\bm x_t$ completely defines the cubic smoothing spline solution $\hat m(t)$.
% {\bf (Did you enforce the $c_1=c_T=0$ constraint?)}

With a positive inverted temperature $\kappa$, an emulated state space model is one such that whose likelihood of $\bm x_T$ conditioned on $y_1,\dots, y_T$ is $\pi(\bm x_T\mid \bm y_T) \propto e^{-\kappa f(\bm x_T)}$, with $f(\cdot)$ defined in (\ref{eq:smoothingspline-objective-ssm}). One possible way to decompose $\pi(\bm x_T\mid\bm y_T)$ into the likelihood of a state space model is the following.
\begin{align}
    \pi(\bm x_T\mid \bm y_T) &\propto \exp\left(-\kappa f(\bm x_T)\right)\nonumber\\
                &=\exp\left(-\kappa\sum_{t=1}^T(y_t - a_t)^2 - \dfrac{4\lambda\kappa}{3}(\sum_{t=1}^{T-1}(c_t^2+c_tc_{t+1}+c_{t+1}^2)\right)\nonumber\\
                % &=\left(\prod_{t=1}^Te^{-\kappa (y_t - a_t)^2}\right)\exp\left(-\dfrac{2\lambda\kappa}{3(2-\sqrt{3})}\sum_{t=1}^{T-1}(c_{t+1}+(2-\sqrt{3})c_t)^2 \right)\nonumber\\
                &=\left(\prod_{t=1}^Te^{-\kappa (y_t - a_t)^2}\right)
                \left(\prod_{t=2}^T e^{-\dfrac{2\lambda\kappa}{3(2-\sqrt{3})}(c_t+(2-\sqrt{3})c_{t-1})^2}\right),\label{eq:ss-likelihood}
\end{align}
where $\kappa$, the ``temperature" parameter, controls the shape of the distribution.

The second term of (\ref{eq:ss-likelihood}) provides a construction of a first order vector auto-regressive process on $\{x_t=(a_t, b_t, c_t)\}_{t=1,\dots, T}$ as the state equation
\begin{align}
    \begin{bmatrix}
    a_t\\
    b_t\\
    c_t
    \end{bmatrix} &= 
    \begin{bmatrix}
    1 & 1 & \sqrt{3}/3\\
    0 & 1 & \sqrt{3} - 1\\
    0 & 0 & - (2-\sqrt{3})
    \end{bmatrix}
    \begin{bmatrix}
    a_{t-1}\\
    b_{t-1}\\
    c_{t-1}
    \end{bmatrix} + 
    \begin{bmatrix}
    1/3\\
    1\\
    1
    \end{bmatrix}\eta_t,\label{eq:ss-state-dynamics}
\end{align}
with $\eta_t \sim \mathcal N(0, \sigma_b^2), \quad\sigma_b^2 = \dfrac{3(2-\sqrt{3})}{4\lambda\kappa}$.
The first term of (\ref{eq:ss-likelihood}) provides the construction of the observation equation
\begin{align}
    y_t &= \begin{bmatrix}
    1 & 0 & 0
    \end{bmatrix}\begin{bmatrix}
    a_t \\
    b_t \\
    c_t 
    \end{bmatrix} + \varepsilon_t,\label{eq:ss-obs-dynamics}
\end{align}
with $\varepsilon_t\sim\mathcal N(0, \sigma_y^2), \quad\sigma_y^2 = 1/(2\kappa)$, and the initial values
\begin{align*}
    a_1&\sim \mathcal N(y_1, \sigma_y^2),\ b_1\sim 1\text{ and } c_1=0.
\end{align*}

\subsubsection{Regularized Linear Regression}\label{sec:emulation-example-lasso}
LASSO \citep{tibshirani1996regression} is a widely-used regularized linear regression estimation procedure that can perform variable selection and parameter estimation at the same time. \\

Consider the regression model
$$\bm Y = \sum_{j=1}^p \beta_j\bm Z_j + \bm \eta$$
where $\bm Z_1,\dots, \bm Z_p\in\mathbb R^n$ are the $p$ covariates that are used to model the dependent variable $\bm Y\in\mathbb R^n$ and $\bm\eta\sim \mathcal N(0, \sigma_y^2I_n)$. A LASSO estimator of $(\beta_1, \dots, \beta_p)$ is the minimizer of 
\begin{equation}
    f(\beta_1,\dots, \beta_p) = \|\bm Y - \beta_1\bm Z_1 - \cdots - \beta_p\bm Z_p\|_2^2 + \lambda \sum_{j=1}^p |\beta_j|.\label{eq:lasso-objective}
\end{equation}
For a fixed set of $(\beta_1,\dots, \beta_p)$, for $t=1,\dots, p$, define the partial residual $\bm \epsilon_t$ as 
\begin{equation}
    \bm \epsilon_t = \bm Y - \sum_{l=1}^t\beta_l\bm Z_l\label{eq:epsilon-def}
\end{equation}
and $\bm \epsilon_0=\bm Y$.
% To discover the sequential structure embedded in (\ref{eq:lasso-objective}), let $\bm \epsilon_j = \bm Y - \sum_{l=1}^j\beta_l\bm Z_l$ for $j=1,\dots, p$ and $\bm\epsilon_0 = \bm Y$. 
Since
$$\|\bm\epsilon_t\|_2^2 = \|\bm\epsilon_{t-1} -\beta_t\bm Z_t\|_2^2 = \|\bm\epsilon_{t-1}\|_2^2 + \|\bm Z_t\|_2^2\left(\beta_t - \dfrac{\bm\epsilon_{t-1}'\bm Z_t}{\|\bm X_t\|_2^2}\right)^2-\dfrac{\left(\bm\epsilon_{t-1}'\bm X_t\right)^2}{\|\bm Z_j\|_2^2},$$
we have
\begin{equation}
    f(\beta_1,\dots, \beta_p) =\|\bm \epsilon_p\|_2^2+\lambda\sum_{t=1}^p|\beta_t|= \|\bm Y\|_2^2 +\sum_{t=1}^p\left\{ \|\bm Z_t\|_2^2\left(\beta_t - \dfrac{\bm\epsilon_{t-1}'\bm Z_t}{\|\bm Z_t\|_2^2}\right)^2-\dfrac{\left(\bm\epsilon_{t-1}'\bm Z_t\right)^2}{\|\bm Z_t\|_2^2}+\lambda |\beta_t|\right\}.\label{eq:lasso-sequential}
\end{equation}
Let $x_t=\beta_t$ and $\bm x_t=(\beta_1,\dots, \beta_t)$. An emulated state space model can be designed so that
\begin{align}
    \pi(\bm x_p) &\propto \exp\left\{-\kappa f(\bm x_p)\right\}
    \propto\prod_{t=1}^p\exp\left\{-\kappa \|\bm Z_t\|_2^2\left(x_t - \dfrac{\bm \epsilon_{t-1}'\bm Z_t}{\|\bm Z_t\|_2^2}\right)^2\right\}\times \prod_{t=1}^p \exp\left\{-\kappa\lambda|x_t| +\kappa \dfrac{(\bm\epsilon_{t-1}'\bm Z_t)^2}{\|\bm Z_t\|_2^2}\right\}.\label{eq:lasso-likelihood}
\end{align}
The first term of (\ref{eq:lasso-likelihood}) leads to the state equation
\begin{equation}
    p_t(x_t\mid \bm x_{t-1})\propto \exp\left\{-\kappa \|\bm Z_t\|_2^2\left(x_t - \dfrac{\bm \epsilon_{t-1}'\bm Z_t}{\|\bm Z_t\|_2^2}\right)^2\right\},\label{eq:lasso-state}
\end{equation}
and the second term leads to the observation equation
\begin{equation}
    g_t(w_t\mid \bm x_t)\propto \alpha_t\exp\{-\alpha_tw_t\},\label{eq:lasso-obs}
\end{equation}
where 
$$\alpha_t = \exp\left\{-\kappa\lambda|x_t| +\kappa \dfrac{(\bm\epsilon_{t-1}'\bm Z_t)^2}{\|\bm Z_t\|_2^2}\right\},$$
with observation $w_t=0$ for all $t$.

Note that $\bm\epsilon_{t-1}$ is a function of $\bm x_{t-1}$ as defined in (\ref{eq:epsilon-def}) and is available at time $t$.
The observation equation $g_t$ and the observation value $w_t=0$ are imposed to incorporate $\alpha_t$ in $\pi(\bm x_p)$. The emulation for LASSO can be extended to other penalized regression with different penalty terms by changing $\alpha_t$ accordingly.

\subsubsection{Optimal Trading Path}\label{sec:emulation-example-trading}
In asset portfolio management, the optimal trading path problem is a class of optimization problems which typically maximizes certain utility function of the trading path \citep{markowitz1959portfolio}.
\cite{kolm2015multiperiod} and \cite{irie2016bayesian} proposed to turn such an optimization problem to an emulated state space model.
To be more specific, let $\bm x_T=(x_0,\dots, x_T)$ be a trading path in which $x_t$ represents the position held at time $t$. \cite{kolm2015multiperiod} propose to maximize the following utility function. 
\begin{equation}
    u(\bm x_T) = -\sum_{t=1}^T c_t(x_t - x_{t-1}) - \sum_{t=0}^T h_t(y_t - x_t),\label{eq:tp-utility}
\end{equation}
where $(y_0, \dots, y_T)$ is a predetermined optimal trading path in
an ideal world without trading costs, typically obtained by maximizing the risk-adjusted expected
return under the Markowitz mean-variance theory \citep{markowitz1959portfolio}. \cite{kolm2015multiperiod} provides a construction of $(y_0, \dots, y_T)$ based on the term structure of the underlying asset's \emph{alpha} (the excess expected return relative to the market).
% {\bf (Is there a starting value for $y_t$? Maybe add 'xxx provides a detailed construction'.)} 
Let $c_t(\cdot)$ represent the transaction cost which is often assumed to be a quadratic function of the absolute position change $|x_t - x_{t-1}|$. Without loss of generality, we parametrize it as
$$c_t(|x_t - x_{t-1}|) = \dfrac{1}{2\sigma_x^2} \left(|x_t - x_{t-1}|^2 + 2\alpha |x_t - x_{t-1}|+\alpha^2\right),$$
where $\alpha$ is a non-negative constant related to the volatility and liquidity of the asset \citep{kyle2011market}. Let $h_t(\cdot)$ be the utility loss due to the departure of the realized path from the ideal path. We use the squared loss
$$h_t(y_t-x_t) = \dfrac{1}{2\sigma_y^2}(y_t-x_t)^2.$$
Then the objective function is 
$$\pi(\bm x_T\mid \bm y_T)\propto e^{-\kappa u(\bm x_T)} \propto\prod_{t=1}^T\exp\left(-\dfrac{\kappa(|x_t - x_{t-1}| + \alpha)^2}{2\sigma_x^2}\right)\prod_{t=1}^T\exp\left(- \dfrac{\kappa (y_t - x_t)^2}{2\sigma_y^2}\right).$$
Taking the position constraint $x_0 = x_T$ into consideration as discussed in \cite{cai2018resampling}, an emulated state space model can therefore be constructed as
\begin{align}
    p_t(x_t\mid x_{t-1}) &\propto \exp\left(-\dfrac{\kappa(|x_t - x_{t-1}| + \alpha)^2}{2\sigma_x^2}\right),\label{eq:tp-state}\\
    g_t(y_t\mid x_t) & \propto \exp\left(- \dfrac{\kappa (y_t - x_t)^2}{2\sigma_y^2}\right).\label{eq:tp-observation}
\end{align}
With the state equation (\ref{eq:tp-state}) and the observation equation (\ref{eq:tp-observation}), the corresponding state space model has a likelihood function proportional to $\exp(-\kappa u(\bm x_T))$.

\subsubsection{L1 Trend Filtering}\label{sec:emulation-example-l1}
L1 trend filtering \citep{kim2009ell_1} is a variation of Hodrick-Prescott filtering \citep{hodrick1997postwar}. An $\ell$1 trend filtering on $y_1,\dots, y_T$ is defined to be the minimizer of the objective function
\begin{equation}f(x_1,\dots, x_T) = \sum_{t=1}^T (Y_t - x_t)^2 + \lambda \sum_{t=2}^{T-1}|x_{t-1} - 2x_t + x_{t+1}|.\label{eq:objective-l1}
\end{equation}
Minimizing (\ref{eq:objective-l1}) tends to produce a piece-wise linear function due to the $\ell_1$ penalty on second-order difference.
An emulated state space model is designed to have the following Boltzmann likelihood function.
\begin{equation}
    \pi(\bm x_{T})\propto e^{-\kappa f(\bm x_{T})/2} = \prod_{t=1}^T \exp\left\{-\dfrac{\kappa}{2}(y_t - x_t)^2\right\}\prod_{t=3}^T\exp\left\{-\dfrac{\kappa}{2\lambda}|x_t - (2x_{t-1}-x_{t-2})|\right\}.\label{eq:l1-likelihood}
\end{equation}
The first term of (\ref{eq:l1-likelihood}) leads to the observation equation
\begin{equation}
    y_t = x_t + \epsilon_t,\label{eq:l1-obs}
\end{equation}
where $\epsilon_t\sim\mathcal N(0, \sigma_y^2)$ with $\sigma_y^2 = 1/\kappa$.
The second term of (\ref{eq:l1-likelihood}) leads to the following second order auto-regressive process on the states
\begin{equation}
    x_t = 2x_{t-1} - x_{t-2} + \eta_t,\label{eq:l1-state-dynamics}
\end{equation}
where $\eta_t\sim Laplace(0, \lambda_x)$ with $\lambda_x = 2/(\lambda\kappa)$.

\section{Annealed Sequential Monte Carlo}\label{sec:asmc}
\subsection{Sequential Monte Carlo (SMC)}\label{sec:asmc-smc}
The sequential Monte Carlo (SMC) method is a class of sampling methods designed for state space models.  It utilizes the sequential nature of the state space model and draw samples incrementally with sequential importance sampling and resampling (SISR) scheme. A typical SMC approach is demonstrated in Figure \ref{fig:smc}. 
\begin{figure}[!htb]
\begin{center}
\begin{boxedminipage}{6.5in}
\caption{Sequential Monte Carlo (SMC) Algorithm } \label{fig:smc}
\it{
\begin{itemize}
\item Draw $x_1^{(1)}, \dots, x_1^{(n)}$ from $p_1(x_1)$ and initialize all weights $w_0^{(i)} = 1$ for $i = 1,\dots, n$.
\item At times $t=2,\cdots,T$:
\begin{itemize}

\item Propagation: For $i=1,\cdots,n$,
\begin{itemize}
\item Draw $x_t^{(i)}$ from distribution $q_t(x_t\mid \bm x^{(i)}_{t-1})$ and set $\bm x_{t}^{(i)}=(x_{t-1}^{(i)}, x_t^{(i)})$.
\item Update weights by setting
$$
w_t^{(i)}\leftarrow w_{t-1}^{(i)}\cdot \frac{ p_t(x_t^{(i)}\mid  \bm x_{t-1}^{(i)})
g_t(y_t\mid \bm x_{t}^{(i)})}{q_t(x_t^{(i)}\mid \bm x_{t-1}^{(i)})}.
$$
\end{itemize}

\item Resampling (optional):
\begin{itemize}

\item Assign a priority score $\beta_{t}^{(i)}$
to each sample $x^{(i)}_{0:t}$,
$i=1,2,\dots, n$.
\item Draw samples $\{J_1, \dots, J_n\}$ from the set $\{1, \dots, n\}$ with replacement, with probabilities proportional to $\{\beta_{t}^{(i)}\}_{i=1,\dots, n}$.
\item Let $\bm x_{t}^{*(i)}=\bm x_{t}^{(J_i)}$ and $w_{t}^{*(i)}= w_{t}^{(J_i)}/\beta_{t}^{(J_i)}$.
\item Return the new set $\{(\bm x_{t}^{(i)}, w_{t}^{(i)})\}_{i=1,\dots, n}\leftarrow \{(\bm x_{t}^{*(i)}, w_{t}^{*(i)})\}_{i=1,\dots, n}$.
\end{itemize}

\end{itemize}
\item Return the weighted sample set $\{(\bm x_{T}^{(i)}, w_T^{(i)})\}_{i=1,\dots, n}$.
\end{itemize}
}
\end{boxedminipage}
\end{center}
\end{figure}

The function $q_t(\cdot)$ in the propagation step in Figure \ref{fig:smc} is the proposal distribution. As discussed in \cite{LinChenLiu13_delay}, the ``perfect" choice for the proposal is the conditional distribution with full information set such that $q_t(x_t\mid \bm x_{t-1}) = p(x_t \mid \bm x_{t-1}, \bm y_{T})$. However, in most cases, this conditional probability is impossible to evaluate or to sample from at time $t$. The priority score $\beta_t$ is the weight used in the resampling step, which
quantifies the sampler's preference over different sample paths. The most common choice of $\beta_t$ is $\beta_t^{(i)} \propto w_t^{(i)}$.
Different variations of the SMC algorithm choose different proposal distributions and different priority scores. The Bayesian particle filter \citep{gordon1993novel} sets $q_t(x_t\mid \bm x_{t-1}) = p_t(x_t\mid \bm x_{t-1})$. It works well when the observations $y_T$ are relatively noisy compared with the state equation part. With accurate observations, the independent particle filter \citep{lin2005independent} uses $q_t(x_t\mid \bm x_{t-1}) \propto g_t(y_t\mid x_t)$. As an important (with certain additional cost) compromise over the Bayesian particle filter and the independent particle filter, \cite{kong1994sequential} and \cite{liu1998sequential} suggests to adopt $q_t(x_t\mid \bm x_{t-1})\propto p_t(x_t\mid \bm x_{t-1})g_t(y_t\mid x_t)$ to reduce variance. 
% All above methods use the default priority score $\beta_t^{(i)} = w_t^{(i)}$. 
Other sequential Monte Carlo methods focus on finding more appropriate priority scores in resampling with the help of future information. The auxiliary particle filter \citep{pitt1999filtering} conducts resampling with the priority score $\beta_t^{(i)} = w_t^{(i)}p(y_{t+1}\mid x_t)$. The delayed sampling method \citep{chen2000adaptive, LinChenLiu13_delay} looks ahead $\Delta$ steps further and uses $\beta_t^{(i)} = w_t^{(i)} p(y_{t+1},\dots, y_{t+\Delta}\mid x_t)$. 

In emulation for optimization, we are more interested in generating samples in the high probability region of $\pi(\bm x_T)$, hence our problem is essentially a smoothing problem. 
% Further improvement can be made based on the original sequential Monte Carlo algorithm in Figure \ref{fig:smc}. 
\cite{Briers&Doucet&Maskell2010} proposed to use a generalization of two-filter smoothing formula to sample approximately from the joint distribution $\pi(\bm x_T)$. Additional local MCMC moves can be adopted to fight degeneracy \citep{gilks2001following}. Many other SMC smoothing algorithm implementations are proposed to reduce the potential degeneracy in samples. See, for example, \cite{godsill2004monte, del2010forward, Briers&Doucet&Maskell2010, Guarniero2017iterated}

\subsection{Finding the Most Likely Path}\label{sec:mlp}
With emulation, finding the optimum of $f(\bm x)$ is now equivalent to finding the mode, or the most likely state path (MLP), of $\pi(\bm x_T)$,
\begin{equation}
    \bm x^*_{T} = \argmax_{\bm x_{T}\in \mathcal X^T} \pi(\bm x_{T}\mid \bm y_{T}),
\end{equation}
with $\pi(\bm x_T\mid \bm y_T)$ defined in (\ref{eq:posterior}) and $\mathcal X$ being the common support for all latent variables. 
By construction, the mode, which is the optimum of $f(\bm x)$, does not depend on $\kappa$ used in (\ref{eq:boltzmann}). 

In this article, we focus on finding the MLP from Monte Carlo samples. A set of weighted Monte Carlo samples from the distribution $\pi(\bm x_T)$ can be generated by Sequential Monte Carlo (SMC) and its various implementation schemes. 
Let $\{(\bm x_T^{(i)}, w_T^{(i)}\}_{i=1,\dots, n}$ be the samples drawn from the emulated state space model using the SMC algorithm in Figure \ref{fig:smc}.
A natural and easy way is to use the empirical MAP path such that 
\begin{equation}
    \hat{\bm x}_{T}^{(map)} = \argmax_{\bm x_{T} \in \{\bm x^{(i)}_T\}_{i=1,\dots, n}} \pi(\bm x_{T}\mid \bm y_{T}).\label{eq:map}
\end{equation}
Although the empirical MAP involves the least computation given the Monte Carlo samples, it usually requires a very large sample size to achieve high accuracy, especially when the dimension $T$ is large. 

Note that when we use the Boltzmann-like target distributions as in the examples shown above, the MLP is the same under different $\kappa$. However the distribution 
$\pi(\bm x_{T}\mid \bm y_{T}, \kappa)$ is more flat for small $\kappa$ (high temperature) and is more concentrated around the MLP for large $\kappa$. Hence the empirical MAP path tends to be more accurate if the Monte Carlo samples are generated from the target distribution with large $\kappa$.
When $\kappa$ is sufficiently large, the average sample path is also a good estimate of the MAP.
However, it is  much more difficult to generate Monte Carlo samples with large $\kappa$ due to the tendency of being trapped in local optimum. Simulated annealing approach provides a natural bridge to link the high temperature system with easily generated samples with the low temperature system with more accurate estimates. 
% {\bf (Can we add: When $\kappa$ is sufficiently large, the average sample path is also a good estimate of the MAP?)}

\subsection{Annealed SMC}\label{sec: asmc-asmc}
We propose a simulated annealing algorithm for sequential Monte Carlo on state space models. The idea comes from the thermodynamics analogue discussed in the previous section. When the function $\xi(\cdot)$ is chosen to be Boltzmann-like as in (\ref{eq:boltzmann}), the Monte Carlo samples from the emulated state space models correspond to a random sample set from the non-interacting particles in a thermodynamic equilibrium system as discussed in Section \ref{sec:emulation-sse}.
% with the energy levels $f(\cdot)$ and the temperature $1/\kappa$ (the Boltzmann constant $k_B$ is ignored here for simplicity). 
If the temperature cools down to $0$ slowly enough such that the system is approximately in thermodynamic equilibrium for any temperature in between, all particles will condense to the base energy level. The idea of simulated annealing to analogize the physical system was proposed and discussed in \cite{Kirkpatrick1983Optimization}. 

To mimic the thermodynamic procedure, we propose the following system to simulate the annealing procedure for the SMC samples. Let $0<\kappa_0 < \kappa_1 < \dots < \kappa_K$ be an increasing sequence of inverse temperatures. Suppose at $\kappa_0$, a base emulated state space model is constructed as
\begin{equation}
\pi(\bm x_{T};\kappa_0)\propto e^{-\kappa_0 f(\bm x_{T})}\propto p_0(x_0)\prod_{t = 1}^T p_t(x_t\mid \bm x_{t-1})g_t(y_t\mid x_t).\label{eq:base-smc}
\end{equation}
At a higher inverse temperature $\kappa_k$, an emulated state space model can be induced from (\ref{eq:base-smc}) such that
\begin{equation}
    \pi(\bm x_{T};\kappa_k)\propto e^{-\kappa_k f(\bm x_{T})}\propto p_0(x_0;\kappa_k)\prod_{t = 1}^T p_t(x_t\mid \bm x_{t-1};\kappa_k)g_t(y_t\mid x_t;\kappa_k), \label{eq:annealed-smc}
\end{equation}
where 
$$p_t(x_t\mid \bm x_{t-1};\kappa_k)\propto \left[p_t(x_t\mid \bm x_{t-1})\right]^{\kappa_k/\kappa_0}\quad\text{and}\quad g_t(y_t\mid x_t;\kappa_k)\propto\left[g_t(y_t\mid x_t)\right]^{\kappa_k/\kappa_0}$$
are the corresponding state equations and observation equations at $\kappa_k$. 
The starting inverse temperature $\kappa_0$ is usually chosen to be relatively small such that the function $\pi(\bm x_{T};\kappa_0)\propto e^{-\kappa_0 f(\bm x_{T})}$ is relatively flat and is easy to sample from by SMC. We start with $\kappa_0$, draw $\{(\bm x_{0, T}^{(j)}, w_{0,T}^{(j)})\}_{j=1,\dots, m}$ from the base emulated state space model $\pi(\bm x_T;\kappa_0)$. For $k=1,\dots, K$, new samples $\{(\bm x_{k, T}^{(j)}, w_{k,T}^{(j)})\}_{j=1,\dots, m}$ are drawn with respect to the distribution $ \pi(\bm x_T;\kappa_k)$ utilizing samples $\{(\bm x_{k-1, T}^{(j)}, w_{k-1,T}^{(j)})\}_{j=1,\dots, m}$ obtained at $\kappa_{k-1}$. The procedure is depicted in Figure \ref{fig:annealed-smc}. 
The annealed sequential Monte Carlo uses the following proposal distribution at temperature $\kappa_k$:
\begin{equation}
    q_{k,t}(x_t\mid \bm x_{t-1};\kappa_k) \propto \hat p_{k,t}(x_t\mid \bm x_{t-1};\kappa_{k-1}),\label{eq:anneal-proposal}
\end{equation}
where the conditional distribution $\hat p_{k,t}(x_t\mid \bm x_{t-1};\kappa_{k-1})$ is an estimate of $\pi_T(x_t\mid \bm x_{t-1};\kappa_{k-1})$ and can be obtained from the Monte Carlo samples $\{\bm x_{k-1,T}^{(j)}, w_{k-1, T}^{(j)}\}_{j=1,\dots, m}$ under $\kappa_{k-1}$. We will discuss how to obtain such an estimate later. Since $\kappa$ increases slowly, $\pi_T(x_t\mid \bm x_{t-1};\kappa_{k-1})$ and $\pi_T(x_t\mid \bm x_{t-1};\kappa_{k})$ are reasonably close.
% For sufficiently large $\kappa_K$, all the samples $\{(x_0^{[K](i)},\dots, x_T^{[K](i)})\}_{i=1,\dots,n}$ lie in a small neighbor region around the true optimal $x_{0:T}^*$. The optimal path can be obtained by either the Viterbi algorithm or even the sample average. The complete annealed SMC procedure is demonstrated in Figure \ref{fig:annealed-smc}.

\begin{figure}[!htbp]
    \begin{center}
        \begin{boxedminipage}{6.5in}
        \caption{Annealed Sequential Monte Carlo Algorithm}
        \label{fig:annealed-smc}
        \it{
        \begin{itemize}
            \item Draw $\{(\bm x_{0, T}^{(j)}, w_{0,T}^{(j)})\}_{j=1,\dots, m}$ from $\pi(\bm x_T; \kappa_0)$ with SMC in Figure \ref{fig:smc}, using a set of proposal distributions $q_{1,t}(x_t \mid \bm x_{t-1};\kappa_0)$.
            \item For $k = 1,\dots, K$, draw $\{(\bm x_{k, T}^{(j)}, w_{k,T}^{(j)})\}_{j=1,\dots, m}$ from $
            \pi(\bm x_T; \kappa_k)$ with SMC in Figure \ref{fig:smc} using the proposal distribution
            $$q_{k,t}(x_t\mid \bm x_{t-1};\kappa_k)\propto\hat p_{k,t}(x_t\mid \bm x_{k, t-1}^{(j)}),$$
            where the right hand side is an estimate of $\pi_T(x_t\mid \bm x_{t-1};\kappa_{k-1})$. 
            \item Obtain an estimate of the  most likely path from $\{(\bm x_{K, T}^{(j)}, w_{K,T}^{(j)})\}_{j=1,\dots, m}$ %by either empirical MAP method, the Viterbi algorithm or the sample average.
        \end{itemize}
        }
        \end{boxedminipage}
    \end{center}
\end{figure}

With a sufficiently large $\kappa_K$, samples from the target distribution $\pi(\bm X_T;\kappa_K)$ are highly concentrated around the true optimal path $\bm x_T^*$ and hence are useful in inferring the most likely path. However, sampling from $\pi(\bm x_{T};\kappa_K)$ directly is usually difficult due to the challenge in finding appropriate proposal distributions, which significantly affects the Monte Carlo sample quality.
Annealed SMC provides an iterative procedure to the difficult sampling problem under $\kappa_K$ by utilizing the samples obtained at higher temperature.
On one hand, annealed SMC provides a relatively ``flat" and easy-sampling starting distribution $\pi(\bm x_T;\kappa_0)$ and designs a slow-changing path connecting $\pi(\bm x_T;\kappa_0)$ to the desired ``sharp" distribution $\pi(\bm x_T;\kappa_K)$. On the other hand, for each iteration $k=1,\dots, K$, annealed SMC adopts an optimal proposal distribution $p(x_t\mid \bm x_{t-1}, \bm y_T;\kappa_{k-1})$, which incorporates the full information set $\bm y_T$ and is usually difficult to evaluate in conventional SMC implementations. In annealed SMC, the proposal distribution is estimated by sample paths from the previous iteration. The details in estimating the proposal distribution will be discussed in Section \ref{sec:asmc-practial}.

The conventional simulated annealing algorithm \citep{Kirkpatrick1983Optimization} is a variation of Markov Chain Monte Carlo (MCMC), which adapts Metropolis-Hastings algorithm \citep{Metropolis1953Equation, Hastings1970Monte} with an extra temperature control. The convergence of the conventional simulated annealing algorithm is given by \cite{Granville1994Simulated}. However, different from the conventional simulated annealing, annealed SMC does not require for a mixing condition as usually shown in MCMC algorithms. At each iteration at $\kappa_k$, the samples are always properly weighted with respect to the target distribution $\pi(\bm x_T;\kappa_k)$ because of the weight adjustments. The convergence of SMC samples is discussed in \cite{crisan2000convergence}.

\subsection{Practical Issues}\label{sec:asmc-practial}
In annealed SMC, at temperature $1/\kappa_k$, we need to estimate the proposal distribution $q_{k,t}(x_t\mid \bm x_{t-1};\kappa_k) = \hat p_{k, t}(x_t\mid \bm x_{t-1})$ with the sample paths from the previous iteration $\{\bm x_{k-1, T}^{(j)}\}_{j=1,\dots, m}$. Notice that, the weighted samples $\{(\bm x_{k-1, T}^{(j)}, w_{k-1, T}^{(j)})\}_{j=1,\dots, m}$ follow the distribution $\pi(\bm x_{t}\mid \bm y_{T};\kappa_{k-1})$. Therefore, estimating the proposal distribution is equivalent to estimating the conditional distribution from a sample set drawn from the joint distribution. Here we mention two methods to sample from such a conditional probability.\\
\textbf{Parametric Approach.}\
For each time $t$, suppose $\{\Psi_{t, \theta}(\cdot)\}$ is a parametric family of distributions defined on $\mathcal X^{t+1}$ and indexed by $\theta$. The joint distribution of $\bm x_{t}$ conditioned on $\bm y_{T}$ under $\kappa_{k-1}$ is approximated by one of the distributions in the family. Specifically, let
$$
\theta^*_{t, k-1} = \argmax_\theta \prod_{i=1}^mw_{k-1, T}^{(i)}\log\psi_{t, \theta}(\bm x_{k-1,t}^{(i)}),
$$
where $\psi_{t, \theta}$ is the corresponding probability density/mass function of $\Psi_{t, \theta}$. Denote the conditional probability induced from $\Psi_{t, \theta}(\bm x_t)$ as $\psi_{t,\theta}(x_t\mid \bm x_{t-1})$. The joint distribution of $\bm x_{t}\mid \bm y_{T}, \kappa_{k-1}$ is approximated by $\psi_{t, \theta^*_{t, k-1}}(\bm x_{t})$ and the proposal distribution $q_t(x_t\mid \bm x_{t-1};\kappa_k)$ is estimated by $\psi_{t, \theta^*_{t, k-1}}(x_t\mid \bm x_{t-1})$. \\
One common choice for the distribution family is the multivariate Gaussian distributions. In this case, 
$$\psi_{t, \bm\mu_{t}, \bm\Sigma_{0:t, 0:t}}(\bm x_{t}) = \mathcal N\left(\bm\mu_{t}, \bm\Sigma_{0:t, 0:t}\right).$$
The optimal parameter can be obtained by sample mean and sample variance such that
\begin{align*}
    \bm\mu^*_{t, k-1} &= \left.\sum_{i=1}^m w_{k-1,T}^{(i)}\bm x_{k-1,t}^{(i)}\middle /\sum_{i=1}^m w_{k-1,T}^{(i)}\right.,\\
    \bm\Sigma^*_{0:t, 0:t, k-1} &= \left.\sum_{i=1}^mw_{k-1,T}^{(i)} \bm x_{k-1, t}^{(i)}\left[\bm x_{k-1,t}^{(i)}\right]'\middle /\sum_{i=1}^m w_{k-1,T}^{(i)}\right..
\end{align*}
Denote
$$
\bm\mu^*_{t, k-1}=\binom{\bm\mu^*_{t-1, k-1}}{\mu^*_{t, k-1}}\quad \text{and}\quad
\bm\Sigma^*_{0:t, 0:t, k-1}=\begin{bmatrix}
\bm\Sigma^*_{0:t-1, 0:t-1, k-1} & \bm\Sigma^*_{0:t-1, t, k-1}\\
\bm\Sigma^*_{t, 0:t-1, k-1} & \Sigma^*_{t, t, k-1}
\end{bmatrix}.
$$
Then the induced conditional probability has the following closed-form:
$$
p(x_t\mid \bm x_{t-1}, \bm y_{T};\kappa_{k-1}) = \mathcal N\left( \mu_{t\mid 0:t-1, k-1}, \Sigma_{t\mid 0:t-1, k-1}\right),
$$
where the parameters are
\begin{align*}
    \mu_{t\mid 0:t-1, k-1}&=\mu^*_{t,k-1} +\bm \Sigma^*_{t, 0:t-1, k-1}\left[\bm \Sigma^*_{0:t-1, 0:t-1, k-1}\right]^{-1}(\bm x_{t-1}-\bm \mu^*_{t-1, k-1}),\\
    \Sigma_{t\mid 0:t-1, k-1}&=\Sigma^*_{t, t, k-1}- \bm \Sigma^*_{t, 0:t-1, k-1}\left[\bm\Sigma^*_{0:t-1, 0:t-1, k-1}\right]^{-1}\bm\Sigma^*_{0:t-1, t, k-1}.
\end{align*}
The results above for multivariate Gaussian distributions can be easily extended to mixture Gaussian distributions, which can approximate most distributions well.\\
\textbf{Nonparametric Approach.}
When there is no appropriate distribution family to describe the joint distribution of $\bm x_{k-1,t}$, one can sample from the conditional distribution $p(x_t\mid \bm x_{t-1}, \bm y_T;\kappa_{k-1})$ of $\{\bm x_{k-1,T}^{(j)}\}_{j=1,\dots, n}$ nonparametrically. Specifically, suppose $\mathcal K_{\bm b_1}(\cdot)$ and $\mathcal K_{b_2}(\cdot)$ are kernel functions for $\bm x_{t-1}$ and $x_t$, respectively, and it is easy to sample from $\mathcal K_{b_2}(\cdot)$. For any given $x_{k, t-1}^{(j)}$, Figure \ref{fig:sample-conditional} depicts the nonparametric approach to draw $x_{k, t}^{(j)}$ from the conditional distribution $p(x_t\mid \bm x_{t-1},\bm y_T;\kappa_{k-1})$ when the samples $\{(\bm x_{k-1,T}^{(i)}, w_{k-1,T}^{(i)})\}_{i=1,\dots, m}$ properly weighted to $\pi(\bm x_T\mid \bm y_T;\kappa_{k-1})$ are available. 
% Given $\bm x_{t-1}$, Figure \ref{fig:sample-conditional} depicts a nonparametric approach to draw a sample of $x_t$ from the conditional distribution $p(x_t\mid \bm x_{t-1},\bm y_T;\kappa_{k-1})$ with the help of samples $\{\bm x_{k-1,T}^{(i)}\}_{i=1,\dots, n}$.

\begin{figure}[!htbp]
\begin{center}
    \begin{boxedminipage}{6.5in}
    \caption{Sample nonparametrically from a Empirical Conditional Distribution}\label{fig:sample-conditional}
    \it{
    For given $\bm x_{k,t-1}^{(j)}$,
        \begin{itemize}
            \item draw $l$ from $\{1,\dots, m\}$ with probabilities proportional to 
            $$P(l=i)\propto w_{k-1,T}^{(i)}\mathcal K_{\bm b_1}(\bm x_{k-1,t-1}^{(i)} - \bm x_{k,t-1}^{(j)}).$$
            \item draw $\varepsilon$ from the density induced by $\mathcal K_{b_2}(\cdot)$.
            \item return $x_{k,t}^{(j)} = x_{k-1,t}^{(l)} + \varepsilon$.
        \end{itemize}
    }
    \end{boxedminipage}
\end{center}
\end{figure}

The parametric approach often requires the state space model to satisfy certain conditions. For example, when both state equations and observation equations are approximately linear and Gaussian, the multivariate Gaussian distribution family can be used to estimate the conditional distributions. The nonparametric approach can deal with general state space models. However, it often costs much more computing power than the parametric approach.

One issue for both approaches is the high dimensionality. 
Unless the system has a short memory, the conditional distribution at time $t$ involves the high dimensional $\bm x_t$ and with potentially increasing dimension of parameter needed or the dimensions of spaces the nonparametric approach need to operate within.
% To sample from the conditional distribution at time $t$, it involves either estimating the parameters of dimension $t$ in the parametric approach or looking for a neighborhood in a $t$ dimension space in the nonparametric approach. When $t$ is large, it is definitely not efficient to consider the problem in a $t$-dimensional context. 
One solution for reducing dimension of the sampling problem is to use a low-dimensional sufficient statistics. Suppose $S(\bm x_{t-1})$ is a low-dimensional sufficient statistic such that $p(x_{t}\mid \bm x_{t-1}, \bm y_{T};\kappa_{k-1}) = p(x_t\mid S(\bm x_{t-1}), \bm y_{T};\kappa_{k-1})$. Both parametric and nonparametric approaches can therefore be conducted on the joint distribution of $(x_t, S(\bm x_{t-1}))$, which is of lower dimension. In a Markovian system, $S(\bm x_{t-1}) = x_{t-1}$ and the problem reduces to sampling from a much simpler distribution. In an auto-regressive system with lag $\delta$, $S(\bm x_{t-1}) = \bm x_{t-\delta:t-1}$, which is a $\delta + 1$-dimensional system. 
Note that since the estimated conditional distribution is used as a proposal distribution, it is often tolerable to use less accurate estimators for computational efficiency. Hence various approximation and dimension reduction tools can be used, including variational Bayes approximations \citep{tzikas2008variational}.
% {\bf (Can we say the following: Note that since the estimated conditional distribution is used as a proposal distribution, it is often tolerable to use less accurate estimators for computational efficiency. Hence various approximation and dimension reduction tools can be used, including variational Bayes approximations (references).}

Another issue in estimating the conditional distribution from sequential Monte Carlo samples is the sample degeneracy. In SMC, degeneracy refers to the phenomenon that the number of distinct values for some states such as $X_1$ can be  less than the number of Monte Carlo samples, if resampling steps are engaged. The degeneracy problem is crucial for both approaches in sampling from the conditional distribution. 
Therefore, at $\kappa>\kappa_0$, we suggest to conduct resampling only when all propagation steps are finished to prevent the samples from trapping into local maximums.
When high degeneracy is persistent, we suggest to use post-MCMC steps \citep{gilks2001following} to regenerate the samples.
If the system is reversible and SMC can be implemented backward in $t$, alternating forward and backward sampling through the annealing iterations may also reduce the degeneracy problem as it starts with more diversified samples in each temperature iteration.

\subsection{Path refinement with Viterbi algorithm}
A more accurate estimate of the mode can be obtained by using Viterbi algorithm \citep{viterbi1967error} on the discrete space consisting of the SMC samples. Viterbi algorithm is a dynamic programming algorithm originally used to solve the MLP problem in hidden Markov models, where the hidden states are finite. 
Let $\mathcal A_t=\{a_t^{(j)}\}_{j=1,\dots,m}$ be the grid points for $x_t$ and $\Omega=\mathcal A_1\times\dots\times \mathcal A_T$ be the Cartesian product of the grid point sets.
In state space models, the Viterbi algorithm searches for the maximum over all possible combinations of the grid points in $\Omega$. Specifically, the MLP obtained by the Viterbi algorithm is
\begin{equation}
    \hat {\bm x_{T}}^{(viterbi)} = \argmax_{\bm x_T\in \Omega} \pi(\bm x_T\mid \bm y_{T}).\label{eq:viterbi-opt}
\end{equation}
The Viterbi algorithm for state space models based on the grid points $\{a_1^{(j)}\}_{j=1,\dots, m},\dots, \{a_T^{(j)}\}_{j=1,\dots, m}$ is depicted in Figure \ref{fig:viterbi}.

\begin{figure}[!hbtp]
\begin{center}
\begin{boxedminipage}{6.5in}
\caption{Viterbi Algorithm for Markovian State Space Models } \label{fig:viterbi}
\it{
\begin{itemize}
    \item Let $\mathcal A_t=\{a_t^{(j)} \}_{j=1,\dots,m}$ be a set of grid points for $x_t$ for $t=1,\dots, T$.
    \item At time 1, initialize $\ell_0^{(j)}=0$ and $\hat {\bm x}_1^{(j)} = a_1^{(j)}$ for $j=1,\dots, m$. 
    \item At each time $t=2,\dots,T$, for $j=1,\dots, m$, set
    \begin{equation}
        \ell_t^{(j)} = \max_{k\in\{1,\dots, m\}}\ \ell_{t-1}^{(k)}p_t(a_t^{(j)}\mid \hat {\bm x}_{t-1}^{(k)})g_t(y_t\mid a_t^{(j)}), \label{eq:viterbi-dp}
    \end{equation}
    and set $\hat {\bm x}_{t}^{(i)} = (\hat {\bm x}_{t-1}^{(k_j^*)}, a_t^{(j)})$, where $j_j^*$ is the optimal point of (\ref{eq:viterbi-dp}).
    \item Let 
    $$j^* = \argmax_{j\in\{1,\dots, m\}} \ell_T^{(j)}.$$
    return $\hat {\bm x}_{T}^{(j^*)}$.
\end{itemize}
}
\end{boxedminipage}
\end{center}
\end{figure}

The SMC samples drawn from the emulated state space model provides a set of grid points for the Viterbi algorithm. For example, one can set $\mathcal A_t = \{x_t^{(i)}\}_{i=1,\dots,m}$ such that $\Omega = \{x_1^{(i)}\}_{i=1,\dots, m}\times\dots\times\{x_T^{(i)}\}_{i=1,\dots, m}$ is the joint set of all SMC sample points. One can also add and remove grids points to expand coverage with more details around the more important state paths. 
%and with SMC weights such that
%$a_t^{(j)}$ is drawn randomly from 
%$$p(a_t^{(j)}=x) \propto \sum_{i=1}^nw_T^{(i)}\mathcal %K_b(x_t^{(i)}-x),$$
%where $\mathcal K_b()$ is a kernel function with bandwidth $b$. In this case, the grid points at $t$, $\mathcal A_t$ is a sample set follows the marginal posterior distribution $\pi(x_t\mid \bm y_T)=\int_{\mathcal X^{T-1}}\pi(\bm x_T\mid\bm y_T)dx_1\dots dx_{t-1}dx_{t+1}\dots dx_T$.

The Viterbi algorithm explores all combination of sample points and results in a better mode estimation compared with the empirical MAP in (\ref{eq:map}).
However, it has its limitations for implementation with state space models.
One limitation is that the Viterbi algorithm only works on Markovian state space models. In addition, it only works with a non-singular state evolution in which the degree of freedom is the same as the state variable dimension. Otherwise, state paths cannot be re-assembled as Viterbi algorithm tries to achieve. For example, in the cubic spline problem, the state evolution is 
singular. Although one can reduce the dimension of the state variable to make the evolution non-singular, the state evolution then becomes non-Markovian. 
%In a state space model with singular state evolutions, only few elements in $\Omega$ have positive probability and the result from Viterbi algorithm is same as the empirical MAP in (\ref{eq:map}).
% {\bf What is a singular state space model? why?}
Another limitation is the requirement for Monte Carlo sample size. 
The Monte Carlo samples induced $\Omega$ provide a discretization of the support $\mathcal X$ for each time $t$. The accuracy of the Viterbi algorithm strongly dependents on the discretization quality, especially when $\mathcal X$ is continuous. In general, the denser the Monte Carlo samples are around the true MLP, the more accurate the Viterbi algorithm solution is. As a result, it often requires a large Monte Carlo sample size to generate better discretization and to achieve high accuracy with Viterbi algorithm. To reduce the path error $\|\hat x_{1:T}^{(viterbi)} - x_{1:T}^*\|$ by half, the Monte Carlo sample size $m$ needs to be doubled, because the discretization size is reduced by half on average with doubled sample size. On the other hand, the computational cost increases quadratically with the sample size $m$. One possible way to improve is to apply Viterbi algorithm iteratively by shrinking to the high value region of last iteration and regenerating grid points there. Similar to iterative grid search, the iterative Viterbi algorithm may result in a sub-optimal solution.
% For example, Figure \ref{fig:suboptimal} plots the true objective function and the grid points from the last iteration is marked with stars. Based on the observed function value that the grid points, the new grid points are more likely to be placed between $2$ and $4$ to explore the high value region. However, the true optimal is at $0.5$.
% \begin{figure}[!hptb]
%     \centering
%     \includegraphics[scale=0.5]{suboptimal.png}
%     \caption{Iterative Viterbi can be suboptimal when grid points misses the true optimal.}
%     \label{fig:suboptimal}
% \end{figure}

\section{Simulation Results}\label{sec:simulation}
\subsection{Cubic Smoothing Spline}\label{sec:simulation-spline}
In this simulation study, we consider the cubic smoothing spline problem in Section \ref{sec:emulation-example-spline}.
The observations are generated by
$$y_t = \sin(9(t-1)/100) + \zeta_t,$$
for $t = 1,\dots, 50$, with $\zeta_t\sim \mathcal N(0, 1/16)$ and we fix $\lambda = 10$ in the objective function (\ref{eq:smoothingspline-objective}).

\begin{figure}[!htpb]
    \centering
    \includegraphics[scale=0.5]{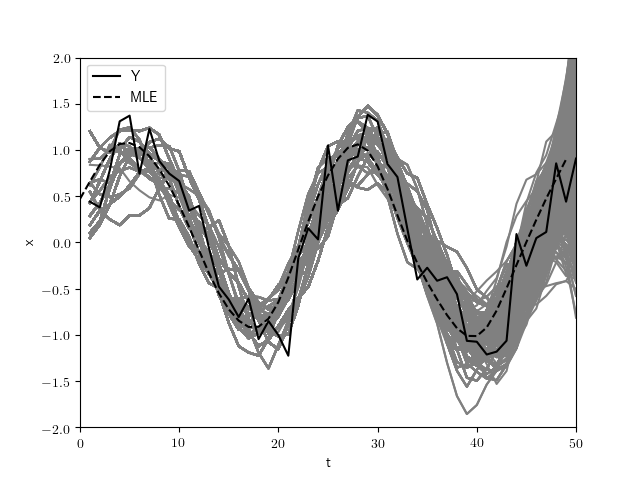}
    \caption{Sample paths at $\kappa_0 = 4$.}
    \label{fig:ss_sample_0}
\end{figure}

Since the dynamic system is linear and Gaussian, the most likely path is obtained by Kalman Smoother \citep{kalman1960new}. We use it as the benchmark. 
We start from the initial inverse temperature $\kappa=\kappa_0=4$. Figure \ref{fig:ss_sample_0} demonstrates $m=1000$ samples (in grey) drawn from the target distribution $\pi(\bm x_T\mid \bm y_T;\kappa_0)\propto \left[\pi(\bm x_T\mid \bm y_T)\right]^{\kappa_0}$ by the SMC algorithm described in Figure \ref{fig:smc} along with the observations $\bm y_T$ (the solid line) and the true most likely path (the dashed line). 

The proposal distribution $q_t(\cdot)$ used at $\kappa_0$ is chosen to be proportional to $p_t(x_t\mid \bm x_{t-1}g_t(y_t\mid x_t)$. At each time $t$, $\eta_t$ is drawn from the proposal distribution $q_t(\eta_t\mid a_{t-1}, b_{t-1}, c_{t-1}, y_t)$, which is a Gaussian distribution in this case. Resampling is conducted when the effective sample size (ESS) defined in (\ref{eq:ess}) is less than $0.3 m$. 
\begin{equation}
    ESS = \dfrac{(\sum_{i=1}^m w_t^{(i)})^2 }{\sum_{i=1}^m (w_t^{(i)})^2}.\label{eq:ess}
\end{equation}

To find the most likely path stochastically and numerically, we apply the annealed SMC approach in Figure \ref{fig:annealed-smc} with a predetermined sequence of inverted temperatures $\kappa_k =  1.5^k\kappa_0$ for $k=1,\dots, 16$. The proposal distribution for the anneal SMC is estimated by the parametric approach. Specifically, since the innovation in the state equation is of one dimension, at $\kappa_k$, we only need to generate proposal samples for $c_t$. It is drawn by first fitting $\{(c_{k-1,t}^{(j)}, a_{k-1,t-1}^{(j)}, b_{k-1,t-1}^{(j)}, c_{k-1,t-1}^{(j)})\}_{j=1,\dots, m}$ with a multivariate Gaussian distribution and then sampling from the conditional distribution. To prevent degeneracy, resampling step is only conducted in the end of each annealing SMC iteration and after each iteration, one step of post-MCMC move is conducted to regenerate sample states. The post-MCMC move uses blocked Gibbs sampling \citep{jensen1995blocking}, due to the special structure of
the state dynamic. At each iteration of the Gibbs sampling, $(x_t,x_{t+1},x_{t+2})$ are updated together.

\begin{figure}
    \centering
    \includegraphics[scale=0.4]{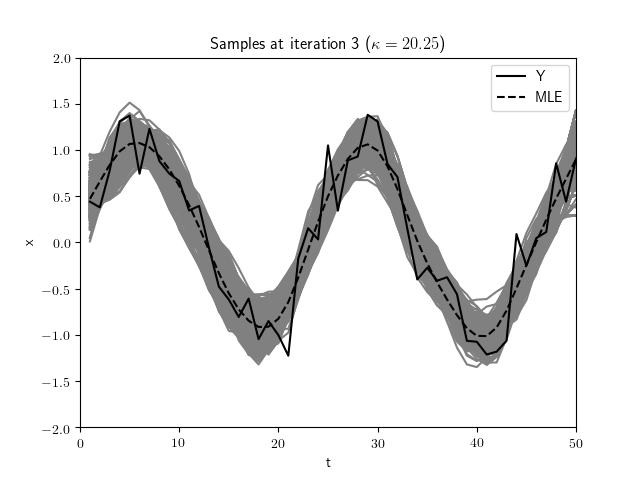}
    \includegraphics[scale=0.4]{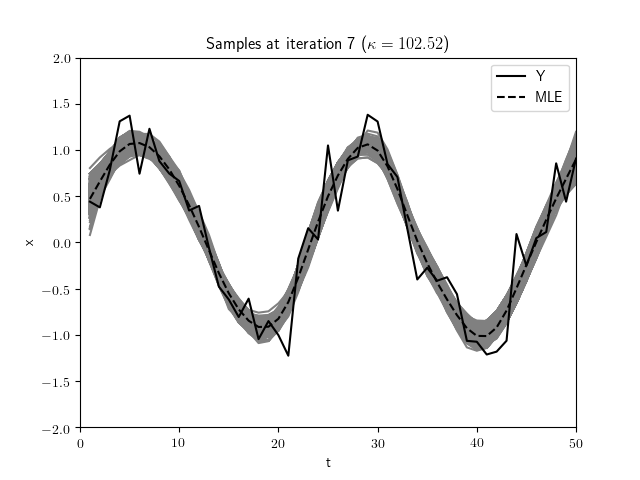}\\
    \includegraphics[scale=0.4]{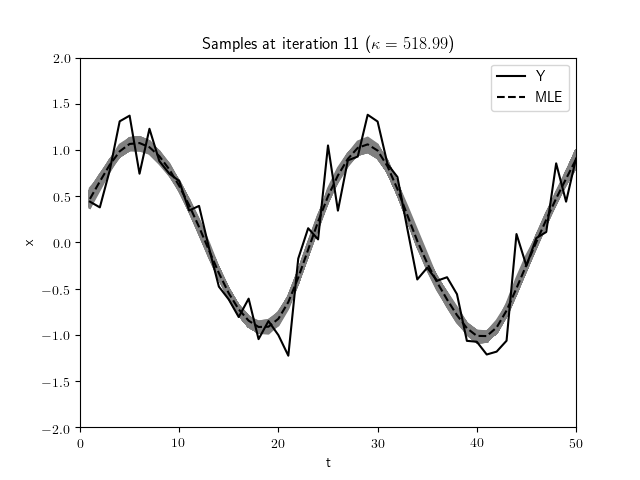}
    \includegraphics[scale=0.4]{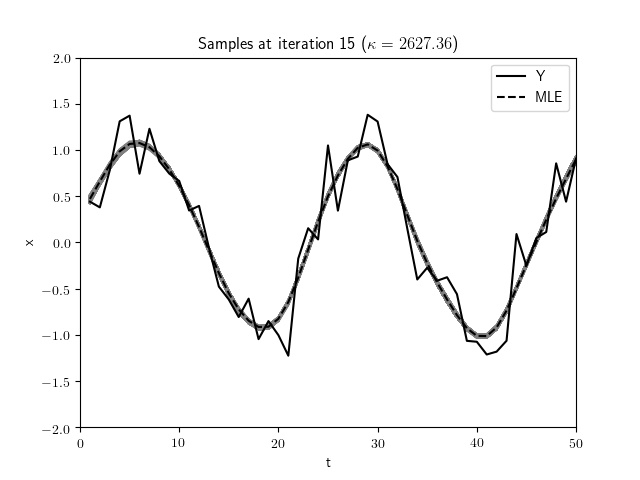}
    \caption{Sample paths at different $\kappa$'s}
    \label{fig:ss_anneal_samples}
\end{figure}

\begin{figure}[!htbp]
    \centering
    \includegraphics[scale=0.5]{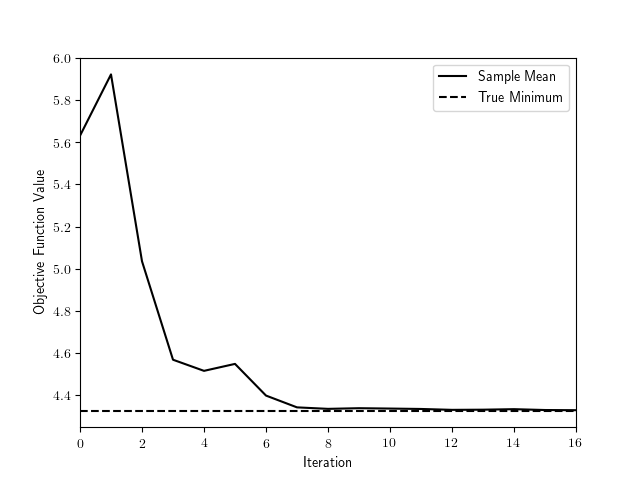}
    \caption{Value of the objective function against the number of iterations}
    \label{fig:ss_convergence}
\end{figure}

Figure \ref{fig:ss_anneal_samples} shows the sample paths (after the post-MCMC step) at the end of different anneal SMC iterations. When the temperature is shrinking to zero as $\kappa$ increases, the sample paths move to a small neighborhood region around the true most likely path. Figure \ref{fig:ss_convergence} shows the value of the objective function at the weighted average path of the samples as for different numbers of iterations. The true optimal value (the objective function value at the optimal path) obtained by the Kalman smoother is plotted as the dashed horizontal line. As the number of iteration increases, the objective function value at the averaged path decreases stochastically and convergences at roughly the 7th iteration. 

To compare the computational efficiency, we record the computing time needed for different approaches, shown in Table \ref{tab:ss_time}. The Scipy approach uses the nonlinear optimizer provided by the python package Scipy \citep{scipy}, which implements the Broyden-Fletcher-Goldfarb-Shanno (BFGS) algorithm by default. The annealed SMC records the time until convergence (the time when the value of objective function is not improved by further iteration). Kalman Smoother is the fastest one due to its deterministic nature in finding the most likely path for linear Gaussian models. Annealed SMC is slower than the nonlinear solver program provided by Scipy, but achieves similar accuracy. We also note that this is a simple convex optimization problem in which a straightforward optimization algorithm such as the Scipy performs well. Our estimation approach is more flexible and this example serves as an illustration of how the algorithm works. 
\begin{table}[!htpb]
    \centering
    \begin{tabular}{|c|c|c|}
    \hline
         Kalman Smoother & Scipy minimizer & Annealed SMC   \\
         \hline
         2.2 ms & 129.6 ms & 232.9 ms\\
         \hline
    \end{tabular}
    \caption{Time spent by different approaches.}
    \label{tab:ss_time}
\end{table}

\subsection{LASSO Regression}\label{sec:simulation-lasso}
In this simulation study, we consider the LASSO regression problem as discussed in Section \ref{sec:emulation-example-lasso}.
We set $n=40$ observations, $p=20$ covariates and $\sigma_y=0.3$. The covariates $( Z_1,\dots,  Z_p)$ are generated from a multivariate normal distribution
$\mathcal N(0, \Sigma)$ where all diagonal elements of $\Sigma$ is 1 and all off-diagonal elements are 0.4. 
% $\lambda=5$ and $\bm Z_1,\dots, \bm Z_p$ are generated from a multivariate normal distribution such that $\bm Z_j$'s have zero means, unit variances and pair-wise covariances of 0.4. 
$\beta$'s are generated i.i.d. according to Bernoulli(0.2). $\lambda$ is set to 5 in the objective function (\ref{eq:lasso-objective}).

\begin{figure}[!htpb]
    \centering
    \includegraphics[scale=0.5]{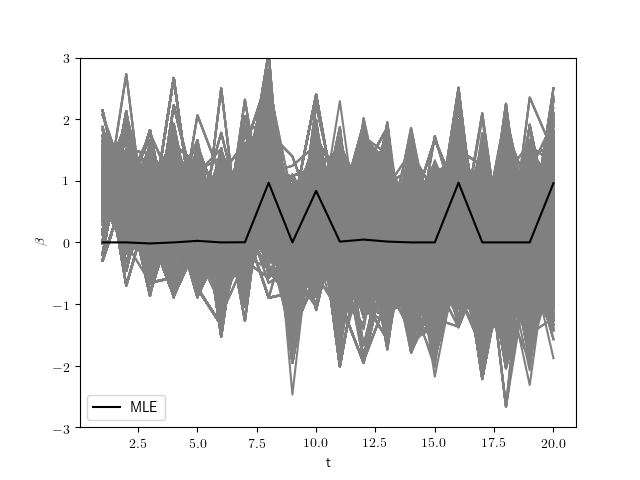}
    \caption{Sample paths at $\kappa_0=0.05$}
    \label{fig:lasso_sample_0}
\end{figure}

% {\bf In LASSO case, did we try to use different random orders in iterations? Would it be better? Do we want (and how) to do final refinement to shrink some coefficients to zero? say in a Vertibi type of algorithm?}
We start from the initial emulated model with the temperature parameter $\kappa =\kappa_0= 0.05$. $m=5000$ samples are drawn from the standard SMC algorithm under the target distribution (\ref{eq:lasso-likelihood}) with $\kappa_0=0.05$. The state equation (\ref{eq:lasso-state}) is used as the proposal distribution and the weight is from the observation equation (\ref{eq:lasso-obs}) as a consequence. Resampling is done when the effective sample size (\ref{eq:ess}) is below $0.3m$. The sampled state paths are plotted in Figure \ref{fig:lasso_sample_0}. 
%The optimal path corresponding to the maximum likelihood estimator under the designed emulating system is obtained by solving the original LASSO problem (\ref{eq:lasso-objective}) using the scikit-learn python package \citep{scikit-learn}. 
The estimated path for solving the original LASSO problem (\ref{eq:lasso-objective}) using the scikit-learn python package \citep{scikit-learn} is treated as the benchmark. . 
% The MLE curve is obtained by a standard LASSO solver.

\begin{figure}[!htpb]
    \centering
    \includegraphics[scale=0.4]{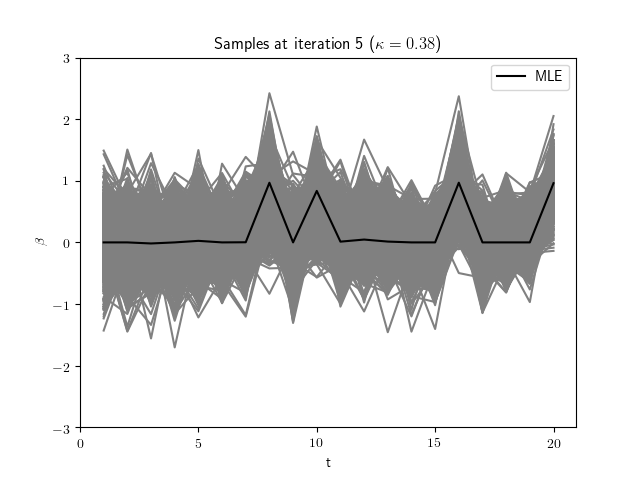}
    \includegraphics[scale=0.4]{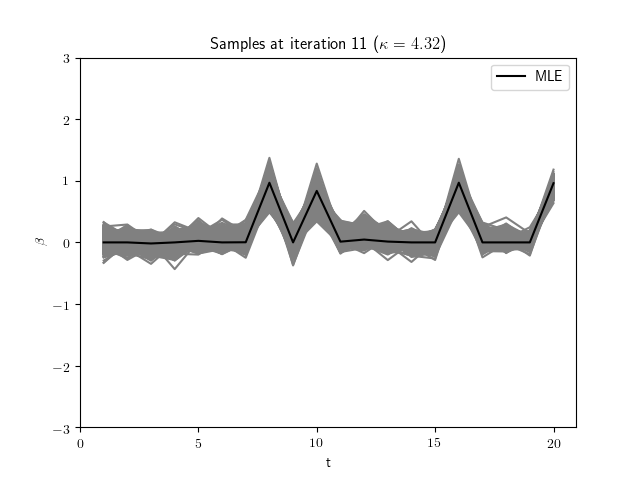}\\
    \includegraphics[scale=0.4]{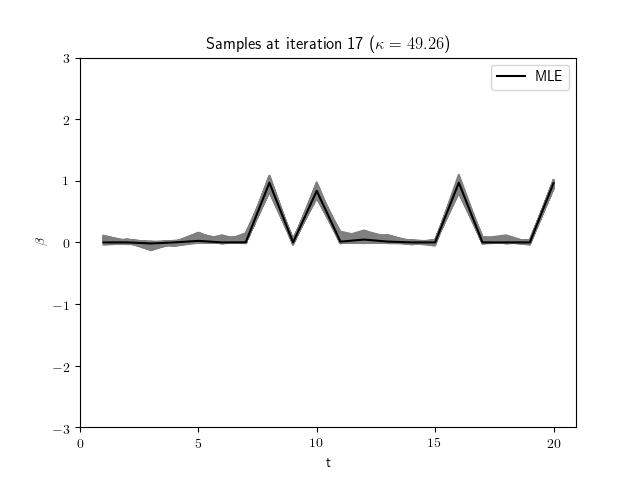}
    \includegraphics[scale=0.4]{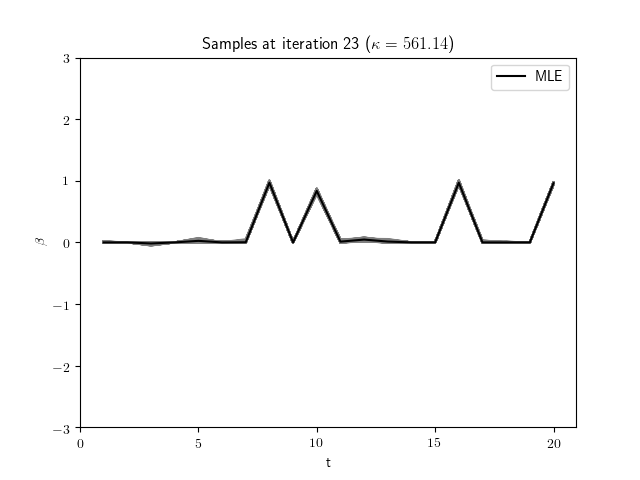}
    \caption{Sample paths at different $\kappa$'s}
    \label{fig:lasso_anneal_samples}
\end{figure}

\begin{figure}[!htpb]
    \centering
    \includegraphics[scale=0.6]{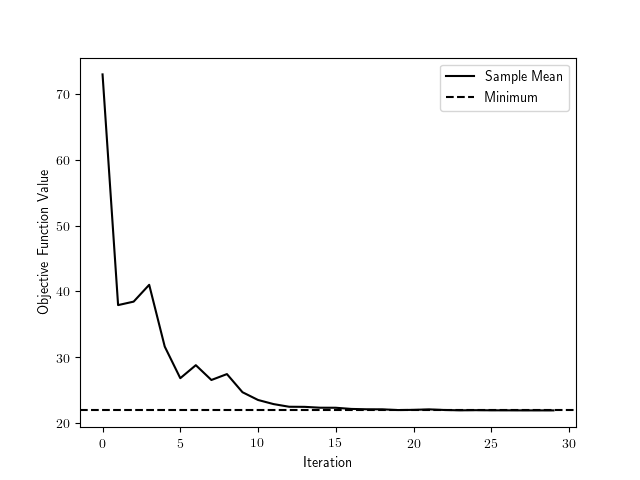}
    \caption{Value of the objective function against the number of iterations}
    \label{fig:lasso_convergence}
\end{figure}

In the subsequent annealing procedure, we use $m=2000$ samples and set $\kappa_k = 1.5^k\kappa_0$ for $k=1,\dots, 30$. The proposal distribution used in the annealing procedure is estimated with a multivariate normal approximation of the joint distribution of $(\beta_{k-1,t},\dots,\dots, \beta_{k-1,1})$. Resampling is done only at the end of each iteration and 10 steps of post-MCMC runs are applied. The post-MCMC runs use the Gibbs sampling approach with the Metropolis-Hasting transition kernel \citep{Metropolis1953Equation, Hastings1970Monte}, where for $t=1,\dots, T$ and for $i=1,\dots, m$, a new value for $\beta_t$ is proposed such that $\tilde \beta_t^{(i)} = \beta_t^{(i)} + \mathcal N(0, \tau^2)$, where $\tau^2 \propto 1/\kappa$, and the proposed move is accepted with the probability $\min(1, \pi(\tilde{\bm x_t}^{(i)}\mid \bm y_T;\kappa)/ \pi(\bm x_T^{(i)}\mid\bm y_T;\kappa))$ with $\tilde{\bm x}_t^{(i)}=(\bm x_{t-1}^{(i)}, \tilde x_t^{(i)}, x_{t+1}^{(i)},\dots, x_T^{(i)})$. Figure \ref{fig:lasso_anneal_samples} plots the sample paths at four different levels of $\kappa$'s. Again, it is seen that the procedure is able to  gradually move the sample paths towards the optimal solution. 
Figure \ref{fig:lasso_convergence} shows the convergence of the values of the objective function in (\ref{eq:lasso-objective}) evaluated at the weighted average of the sample paths. 
% objective function to the one obtained by a standard LASSO solver. 
%Figure \ref{fig:lasso_sample_direct} plots 5000 SMC sample paths drawn directly from the emulated state space model at $\kappa=561.44$ without annealing iterations. Compared with the fourth plot in Figure \ref{fig:lasso_anneal_samples}, the sample paths generated directly at $\kappa = 561.44$ leads to a wrong solution due to the high degeneracy.

After around 17 iterations, the weighted mean of the samples generated from the annealed SMC converges. Due to Monte Carlo variations, the sample paths and the average path cannot shrink the coefficients to exactly zero. It is tempting to run the Viterbi algorithm to refine the estimate, with zeros added to the set of allowed values of the state variables. Unfortunately the state space model designed for the LASSO problem is not Markovian hence Viterbi algorithm cannot be used. However, we used an additional refinement step by iteratively and greedily comparing each estimated state $\hat{x}_t$ (using the average sample path) with zero under the original objective function. The refinement step (with additional 0.063ms in computing time) moved some of the states to zero, and improved the value of the objective function from 21.90356 to 21.899657. The minimum achieved by the Scikit solver is 21.899645.
%with manually added states $x_t=0$ for each time $t$ can result in an estimate of $\bm \beta$ with some coefficients at exactly 0. 
However, such a refinement is based on the knowledge that the solution of Lasso has exactly zero coefficients, and may not be used in other optimization problems. 
Note that, the emulation system can be easily generalized to other types of regularization on parameters by changing the penalty term in (\ref{eq:lasso-obs}) without much efforts and can be adapted much more complex penalty structures.
%\begin{figure}[!htbp]
%    \centering
%    \includegraphics[scale=0.5]{lasso_direct_sample_23.png}
%    \caption{Sample paths of the emulated model at $\kappa = 561.44$ without annealing.}
%    \label{fig:lasso_sample_direct}
%\end{figure}

\subsection{Optimal Trading Path}\label{sec:simulation-trading}

In this simulation, we consider the optimal trading path problem in Section \ref{sec:emulation-example-trading}.
Following \cite{cai2018resampling}, we set $T = 20$, $\sigma_x^2 = 0.25$, $\sigma_y^2 = 1$ and $\alpha = 0.5$. The ideal trading path is given by
$$y_t=25\exp\{-(t + 1)/8\}-40 \exp\{-(t + 1)/4\}.
$$

\begin{figure}[!htpb]
    \centering
    \includegraphics[scale=0.6]{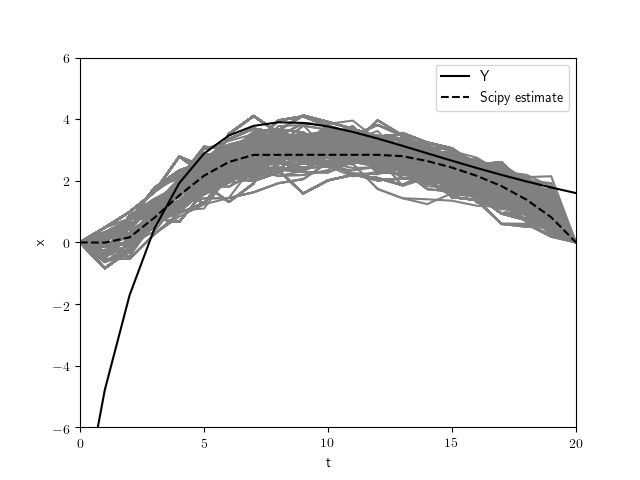}
    \caption{Sample paths at $\kappa_0$}
    \label{fig:tp_sample_0}
\end{figure}

We start from the initial temperature $\kappa = \kappa_0 = 1.0$. The sample paths at $\kappa_0$ is drawn with the constrained SMC \citep{cai2018resampling}, where the resampling step is conducted with priority scores $\beta_t(\bm x_{t}) \propto \hat p(y_{t+1}, \dots, y_T\mid x_t)$. The priority scores are estimated from a set of backward pilot samples \citep{cai2018resampling}. In this example, we use $m^*=300$ backward pilot samples. The resulting $m=1000$ (forward) sample paths are shown in Figure \ref{fig:tp_sample_0}. The observations $y_1, \dots, y_T$, which represent the ideal optimal
trading strategy without the trading cost, are plotted as the solid line. 
An estimated path, marked by dashed line, is provided by the Scipy nonlinear optimization algorithm.

\begin{figure}[!htpb]
    \centering
    \includegraphics[scale=0.4]{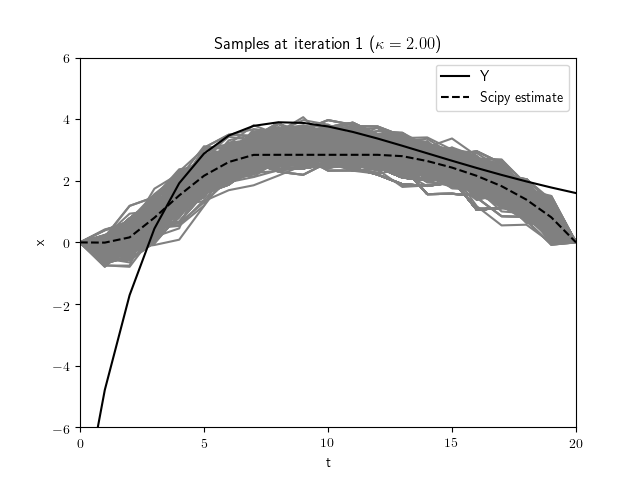}
    \includegraphics[scale=0.4]{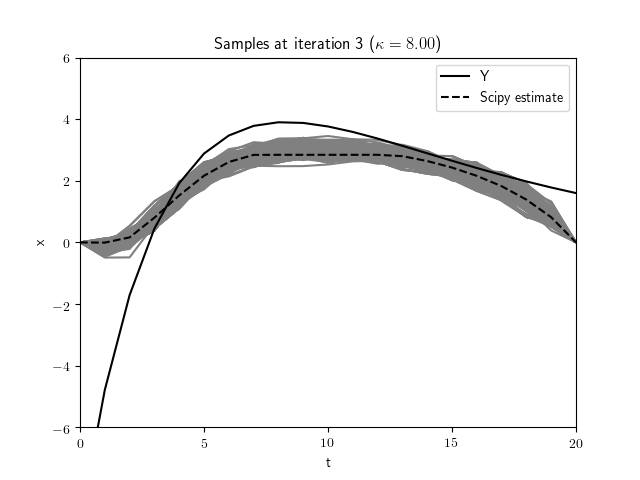}\\
    \includegraphics[scale=0.4]{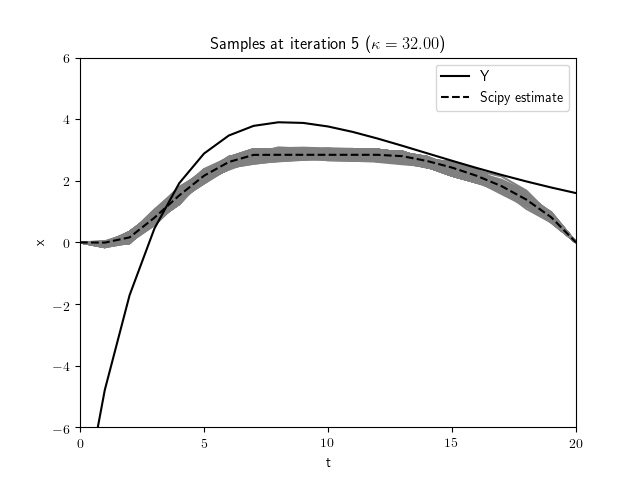}
    \includegraphics[scale=0.4]{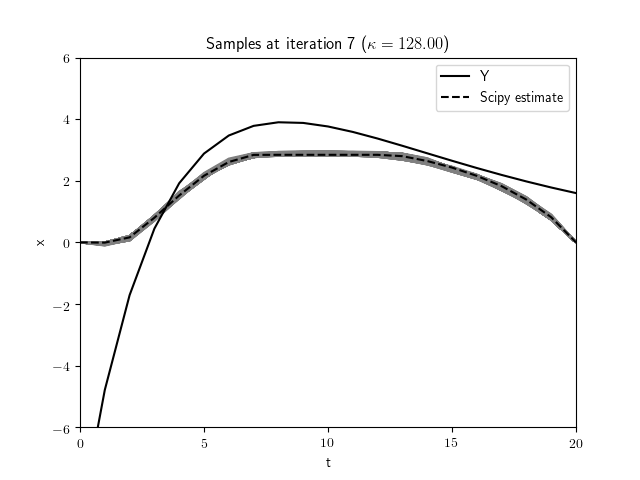}
    \caption{Sample paths at different $\kappa$'s}
    \label{fig:tp_anneal_samples}
\end{figure}

\begin{figure}[!htpb]
    \centering
    \includegraphics[scale=0.6]{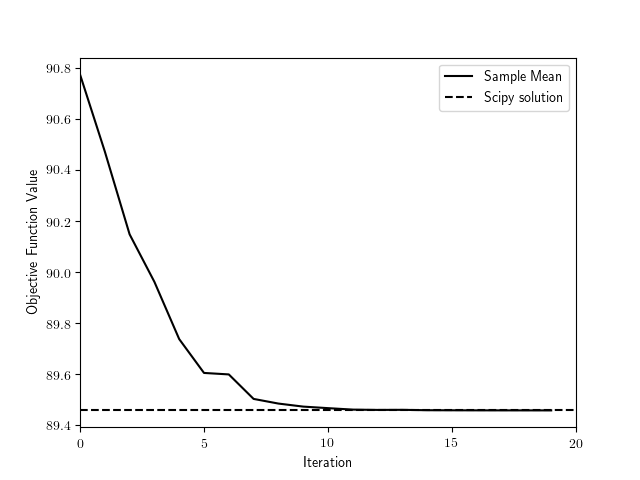}
    \caption{Value of the objective function against the number of iterations}
    \label{fig:tp_convergence}
\end{figure}

We use the following sequence of inverted temperatures for annealing: $\kappa_k = 2^k\kappa_0$ for $k = 1, \dots, 20$. The proposal distribution in the annealed SMC is sampled with the parametric approach by approximating the joint distribution of $x_{k-1,t}$ and $x_{k-1,t-1}$ with a bivariate normal distribution. The annealed $m=1000$ sample paths are resampled at the end of each iteration, and no post-MCMC step is conducted. Samples at several different inverted temperatures are shown in Figure \ref{fig:tp_anneal_samples}. We use the sample average as our estimator for the most likely path. 
% {\bf (But not MAP?)} 
The value of the objective function at the sample average path decreases stochastically as shown in Figure \ref{fig:tp_convergence}. It eventually converges at around the 11th iteration. The optimal objective function value achieved by the annealed SMC is 89.459, while the one obtained by the Scipy nonlinear optimizer is 89.462. The values of the objective function at the sample paths at the 20th iteration has an average of $89.459$ and a standard deviation of $1.09\times 10^{-5}$. The annealed SMC gains some improvement in accuracy at the cost of extra computation. The Scipy nonlinear optimizer takes 78ms while the annealed SMC costs 1.820 seconds for the initial emulated model (including the time of backward sampling) and costs around 2ms for each subsequent iteration. Sampling from the base emulated model costs much more than subsequent iteration for two reasons. First, it requires a large sample size for the base model because of high degeneracy. Second, the end point constraint is imposed and an additional backward pilot run is needed to reduce degeneracy. 


\subsection{L1 Trend Filtering}\label{sec:simulation-l1}
In this simulation study, we consider the $\ell_1$ trend filtering problem in Section \ref{sec:emulation-example-l1}.
We set $T = 60$, $\lambda = 10$ and 
$$y_t = \begin{cases}
\dfrac{t-1}{20} + \mathcal N(0, 0.01),&  1\leqslant t\leqslant 20\\
\dfrac{40-t}{20} + \mathcal N(0, 0.01),& 21\leqslant t\leqslant 40\\
\dfrac{t-41}{20} + \mathcal N(0, 0.01),& 41\leqslant t \leqslant 60.
\end{cases}$$
At $\kappa=\kappa_0=10$, $m=5000$ SMC paths are sampled using the state dynamics (\ref{eq:l1-state-dynamics}) as the proposal distribution. A resampling step is conducted when the effective sample size drops below $0.1m$.
% The samples are plotted in Figure \ref{fig:l1_sample_0}. 
The approximate MLE marked as dashed line is the solution obtained by Scipy nonlinear solver.
The solution shows a piece-wise linear behavior as the $\ell$1 type of penalty appears in the objective function. 

% \begin{figure}[!htpb]
%     \centering
%     \includegraphics[scale=0.6]{l1_sample_0.png}
%     \caption{Sample paths at $\kappa_0=10$.}
%     \label{fig:l1_sample_0}
% \end{figure}

\begin{figure}[!htpb]
    \centering
    \includegraphics[scale=0.4]{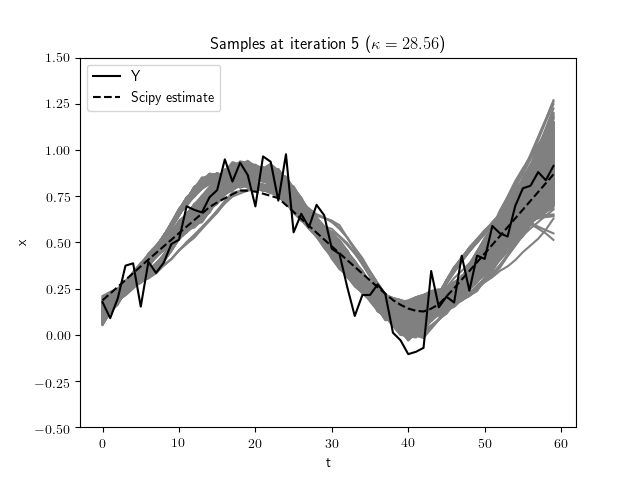}
    \includegraphics[scale=0.4]{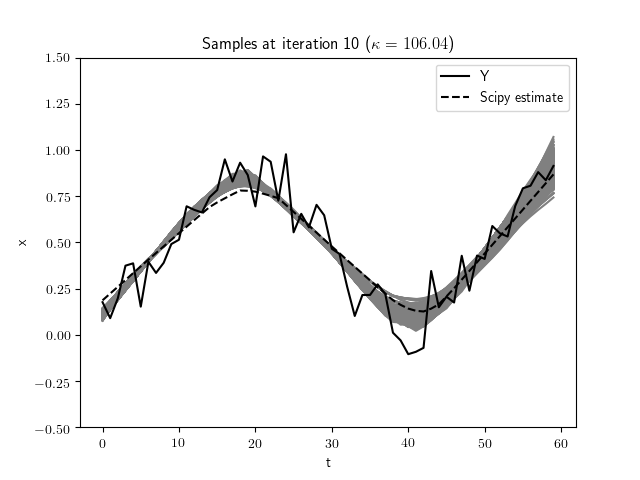}\\
    \includegraphics[scale=0.4]{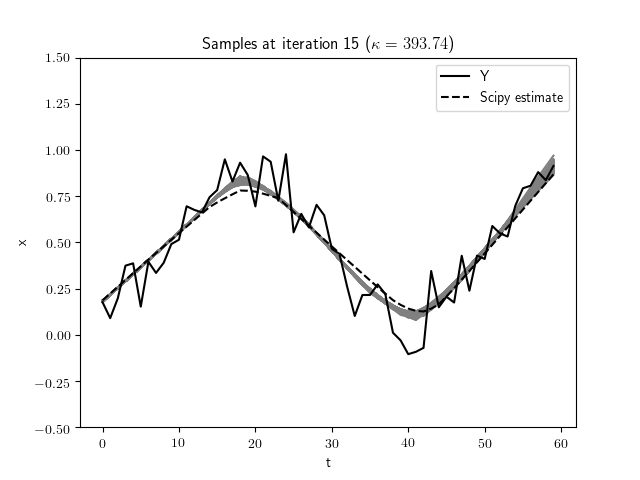}
    \includegraphics[scale=0.4]{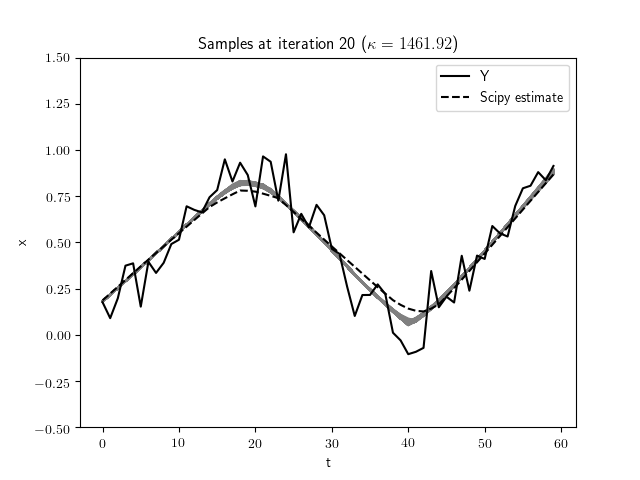}
    \caption{Sample paths at different $\kappa$'s}
    \label{fig:l1_anneal_samples}
\end{figure}

\begin{figure}[!htpb]
    \centering
    \includegraphics[scale=0.6]{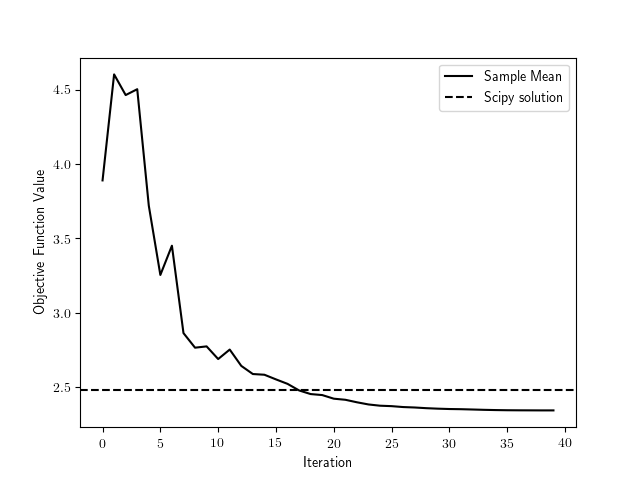}
    \caption{Value of the objective function against the number of iterations}
    \label{fig:l1_convergence}
\end{figure}

We use the following designed annealing sequence $\kappa_k = 1.3^k\kappa_0$ for $k = 1, \dots, 40$ and use $m=2000$ samples for annealing. In each annealing iteration, the proposal distribution used is 
$Laplace(\hat E[x_t\mid x_{t-1}, x_{t-2};\kappa_k], \hat V[x_t\mid x_{t-1}, x_{t-2};\kappa_k]^{1/2}/\sqrt{2})$ where $\hat E$ and $\hat V$ are estimated from the samples from the last iteration $\{(x_{k-1,t}^{(j)}, x_{k-1,t-1}^{(j)}, x_{k-1,t-2}^{(j)})\}_{j=1,\dots,m}$.
% is estimated with the following two steps. In the first step, the samples from the last iteration $\{(x_{k-1,t}^{(j)}, x_{k-1,t-1}^{(j)}, x_{k-1,t-2}^{(j)})\}_{j=1,\dots,m}$ is approximated with a multivariate normal such that the conditional mean and variance can be estimated. Denote the estimated conditional mean and variance as $\hat E[x_t\mid x_{t-1}, x_{t-2};\kappa_k]$ and $\hat V[x_t\mid x_{t-1}, x_{t-2};\kappa_k]$. In the second step, the new proposal samples of $x_t$ at $\kappa_k$ is drawn from $Laplace(\hat E[x_t\mid x_{t-1}, x_{t-2};\kappa_k], \hat V[x_t\mid x_{t-1}, x_{t-2};\kappa_k]^{1/2}/\sqrt{2})$. 
The Laplace distribution has a heavier tail than the normal distribution with the same variance. We found it more efficient to sample from the Laplace distribution to reduce sample degeneracy in this problem. The resampling step is conducted at the end of each iteration and is followed by 10 steps of post-MCMC moves. The post-MCMC steps follow the standard Gibbs sampling as in the LASSO example.
Sample paths at four different $\kappa$'s are displayed in Figure \ref{fig:l1_anneal_samples}. Note that when $\kappa\approx 1462$, the sample paths are different from the nonlinear solver's solution at $t\in[38, 42]$. The value of the objective function at the sample average path shown in Figure \ref{fig:l1_convergence} show that annealed SMC can obtained a smaller objective function value than the Scipy optimizer. The Scipy nonlinear optimizer takes 155ms while annealed SMC costs 22 ms for SMC sampling from the initial emulated model and costs around 160 ms for each subsequent annealing iteration including the post-MCMC runs.
% It reveals that for this kind of high dimensional optimization problem, a general optimizer may not guarantee an accurate result and annealed SMC can perform better in terms of the optimal value.

\section{Summary and Discussion}\label{sec:conclusion}
In this article, we propose a general framework of state space model emulation for high dimensional optimization problems. We demonstrated that, by constructing a proper state space model, many high dimensional optimization problems can be turned into the problem of finding the optimal (most likely) path under the state space model.
And we propose an novel annealed sequential Monte Carlo method to solve the most likely path problem numerically through a simulated annealing scheme with SMC samples. We demonstrate the procedure of state space model emulation with four conventional problems and show how they can be solved using the proposed annealed SMC approach.

The proposed annealed SMC approach shares some similar properties with the traditional simulated annealing methods. Both can optimize a wide range of objective functions including non-convex functions and multi-modal functions, and both often require heavier computation cost than the simpler standard optimization algorithms such as the gradient descent algorithms. However, the annealed SMC approach for state space models is different to the traditional simulated annealing methods in association with MCMC for stochastic optimization
in the following ways. First, emulating an optimization problem into a state space model has its advantage in many problems, especially when the problem is of high dimensional and when the system is inherently dynamic (such as the trading path problem or the $\ell_1$ trend filtering problem) or when the parameters to be estimated inherently play similar roles in the problem (such as the parameters in the regression problem). Second, SMC as an alternative to MCMC has certain advantages in many fixed dimensional problems such as in the problems when the ``dependence" between the parameters in the emulated target distribution is local and (locally) very strong. In these problems, MCMC encounters slow mixing difficulties while SMC naturally takes advantage of such properties. Third, given any temperature, SMC samples target the equilibrium distribution, while MCMC samples often move towards the target distribution gradually. Hence annealed SMC may tolerate faster cooling schedule. Fourth, the inherited parallel structure of SMC allows faster computation. It also adapts to multi-modal problems better.

% Given any temperature $\kappa$, annealed SMC can generate multiple samples directly from the equilibrium simultaneously, as in simulated annealing a burn-in procedure is often required. As shown in our four examples, annealed SMC can tolerate a much more aggressive temperature control compared with the traditional simulated annealing. In annealed SMC, each iteration can decrease the temperature by 15\% to 50\%. 

The state space model emulation and the annealed SMC provide an alternative way to solve high-dimensional optimization problems. Of course, the approach may not be suitable for all problems, due to its high computational cost and its requirement of certain structures. Nevertheless the approach adds to the high dimensional optimization toolbox a useful method for a wide range of complex problems for which the more traditional method may have difficulties to solve. Although the examples shown in this paper do not demonstrate great improvement of the state space emulation approach over the traditional one, they effectively shown how the approach can be implemented and can be used for other problems.
% As more high-dimensional optimization problems arise from the availability of big data, more applications of state space model emulation are to be discovered.

\newpage
\appendix
\bibliographystyle{apa}
\bibliography{reference}

\end{document}